\documentclass[twocolumn,notitlepage,superscriptaddress,10pt]{revtex4-1}
\usepackage{amsmath}
\usepackage{amsfonts}
\usepackage{graphicx}

\begin{document}

\title{Lifetimes of Confined Optical Phonons and the Shape of a Raman Peak  \\ in Disordered Nanoparticles: II. Numerical Treatment}


\author{Sergei V. Koniakhin}
\affiliation{Institute Pascal, PHOTON-N2, University Clermont Auvergne,  CNRS, 4 Avenue Blaise Pascal, Aubi\`ere Cedex  63178, France}
\affiliation{St. Petersburg Academic University - Nanotechnology Research and Education Centre of the Russian Academy of Sciences, St.Petersburg  194021, Russia}
\author{Oleg I. Utesov}
\email{utiosov@gmail.com}
\affiliation{Petersburg Nuclear Physics Institute NRC ``Kurchatov Institute'', Gatchina 188300, Russia}
\affiliation{Department of Physics, St. Petersburg State University, St.Petersburg  199034, Russia}
\author{Andrey G. Yashenkin}
\affiliation{Petersburg Nuclear Physics Institute NRC ``Kurchatov Institute'', Gatchina 188300, Russia}
\affiliation{Department of Physics, St. Petersburg State University, St.Petersburg  199034, Russia}

\date{\today}

\begin{abstract}
Disorder-induced broadening of optical vibrational eigenmodes in nanoparticles of nonpolar crystals is studied numerically. The methods previously used to treat phonons in defectless particles are adjusted for numerical evaluation of the disordered problem. Imperfections in the form of Gaussian and binary disorders as well as surface irregularities are investigated thoroughly in a wide range of impurity concentrations and disorder strengths.
For dilute and weak point-like impurities the regimes of separated and overlapped phonon levels are obtained and the behavior of the linewidth predicted theoretically is confirmed, the crossover scale falls into the actual range of several nanometers. These notions survive for strong dilute impurities, as well. Regimes and crossovers predicted by theory are checked and identified, and minor discrepancies are discussed. To mention a few of them:
slower than in theory increasing of the linewidth with the phonon quantum number for weak disorder and only qualitative agreement between theory and numerics for resonant broadening in strong dilute disorder. The novel phenomena discovered numerically are: ``mesoscopic smearing'' of distribution function in the ensemble of identical disordered particles, inflection of the linewidth dependence on the impurity concentration for light ``dense'' binary impurities, and position dependent capability of strong impurity to catch the phonon. It is shown that surface irregularities contribute to the phonon linewidth less than the volume disorder, and their rate reveal faster decay with increasing of the particle size. It is argued that the results of present research are applicable also for quantum dots and short quantum wires.
\end{abstract}

\maketitle


\section{Introduction}

Manufacturing, characterization, investigation and application of diverse small particles and, particularly, of crystalline dielectric and semiconducting nanoparticles are among the most important and intensively developing areas of contemporary scientific research and technology. The role of nanoparticle studies in domains of optics, quantum computing, chemistry, and material science permanently grows \cite{behler2009nanodiamond,kidalov2009thermal,xia2013nanoparticles,veldhorst2014addressable,Chen2017,andrich2017long,riedel2017deterministic,lin2018catalysis,pichot2019nanodiamond}; various  applications penetrate  biology and medicine, the nanoparticles being used as dyes, carriers, and imaging systems \cite{faklaris2009photoluminescent,walling2009quantum,park2009biodegradable,arnault2015surface,kim2016ultrasmall,kurdyukov2019fabrication}. This imposes the strict necessity for tools and methods of their physical and chemical characterization \cite{mourdikoudis2018characterization,calvaresi2020route}, the particle size is important parameter to know. Such experimental techniques as atomic force microscopy, dynamic light scattering, transmission electron microscopy, calorimetry, X-rays diffraction, Raman spectroscopy, etc., have been used for these purposes \cite{yoshikawa1993raman,yoshikawa1995raman,shenderova2011nitrogen,korobov2013improving,stehlik2015size,stehlik2016high,koniakhin2018ultracentrifugation,dideikin2017rehybridization,chang2017size,bahariqushchi2017correlation,kwon2020evolution,koniakhin2020evidence,kovavrik2020particle}.

Raman spectroscopy is sensitive to the finite size quantization of wavevectors of optical phonons taking place in particles. As a result, the Raman peak for a particle is red shifted comparing to the bulk material~\cite{richter1981one}. Since the introduction of the phonon confinement model (PCM) it became possible to connect this shift with the particle size and therefore to adjust Raman spectroscopy for size probing \cite{campbell1986effects}. However, the PCM approach remains completely phenomenological. Moreover, for smallest nanoparticles it produces very inaccurate  results~\cite{koniakhin2018raman,utesov2018raman,zi1997comparison,meilakhs2016new,gao2019determination}. Numerous efforts undertaken to modify the PCM and to enhance the quality of analysis of the experiment \cite{mochalin2012properties,korepanov2017carbon,korepanov2017quantum,korepanov2020localized,ke2011effect,stehlik2015size,faraci2006modified} did not change this situation principally.

Recently, two mutually related novel descriptions~\cite{koniakhin2018raman,utesov2018raman} of Raman experiments in nanopowders of nonpolar crystals have been proposed to replace the PCM, both utilizing microscopic approach and therefore both having more solid grounds. Unfortunately, these theories (as well as the PCM) incorporate the linewidths of optical vibrational eigenmodes of a particle as fitting parameters.

The theory developed in Ref.~\cite{our3} and in present paper (hereinafter I and II, respectively) is the microscopic approach to disorder-induced phonon line broadening of optical modes in nanopartices which accomplishes the theory of Raman spectra of nanopowders of nonpolar crystals of Refs.~\cite{koniakhin2018raman} and~\cite{utesov2018raman}. This method is much more precise and well grounded than the PCM. In particular, it allows to extract from Raman data four parameters of nanopowders, such as the mean size of particles $L$, the standard deviation of size distribution function $\delta L$, the particle shape parameterized by the effective faceting number $p$ (elongated particles are not considered), and the strength of intrinsic disorder $S$~\cite{our3} (see also~\cite{ourShort}).

Paper I is devoted to the analytical treatment of this problem. Both spectral weights and linewidths $\Gamma_n$ of eigenmodes are calculated as functions of their quantum numbers $n$, nanoparticle shape, size, and the strength of disorder within the framework of three models, namely for Born impurities, in the smooth random impurity potential, and for strong binary disorder. The results are drastically different for the cases of separated phonon levels and for levels belonging to the continuum.

This paper is the numerical continuation of paper I; however, its outcome is not only the numerical verification of formulas obtained in Ref.~\cite{our3} but also the report of several essentially new results such as the ``mesoscopic smearing'' of distribution function or the dependence of the capability for strong impurity to localize the phonon mode on its location in a particle. Even more important, numerical approach allows to deal with more physical realizations of disorder such as so-called NV (nitrogen + vacancy) centers in diamonds. Furthermore, sophisticated and hardly analytically expressible types of disorder (e.g., surface corrugations) are also investigated.


We adopt general DMM-BPM~\cite{koniakhin2018raman} and EKFG~\cite{utesov2018raman} methods applied previously to study phonons in pure particles for numerical treatment of the disordered problem incorporating the procedure of averaging over disorder configurations into the formalism of Green's functions. We examine numerically (point-like and smooth) Gaussian and (point-like) binary disorder varying impurity concentration $c_{imp}$
and atomic mass variation $\delta m /m$ in wide intervals but mostly focusing on dilute regime $c_{imp} \lesssim 0.1 $. For surface corrugations we introduce two models of disorder which we call ``peeled apples model'' and ``nibbled apples model'' which allows to investigate (possible) scaling properties of the disorder.

The analytical predictions made for weak point-like impurities about phonon linewidth dependencies in the form $\Gamma_n \propto \sqrt{S}/L^{3/2}$ for separated phonon levels and  $\Gamma_n \propto S/L$ for overlapped ones are confirmed numerically, and the spatial scale of crossover between these regimes is estimated as lying within the nanometer range. We find that the phonon linewidth indeed growths with its quantum number but slower that it is predicted by the theory. Investigating smooth random potential we observe numerically significant diminishing of $\Gamma (\sigma)$ that occurs at characteristic disorder scale $\sigma \simeq L/2\pi$. We discover numerically and explain the phenomenon of broadening of the distribution function for the ensemble of identical disordered particles which we call ``mesoscopic smearing''.

Next, examining numerically phonons subject to strong dilute disorder we find that the notions of separated and overlapped regimes with their specific $c_{imp}$- and $L$- dependencies for the phonon linewidths survive in this case, either. For dense and very light binary impurities we detected crossover to the novel regime $\Gamma_n \propto c^{3/2}_{imp} /\sqrt{L}$ which originates from multi-impurity scattering processes and proximity to percolation transition. Inspecting the resonant impurity scattering and the formation of optical phonon-impurity (localized) bound states we report a good qualitative agreement between the results of numerical experiment and the theory. The new phenomenon seen in numerics is the rapid decay with the distance from the center of a particle observed for the ability of strong impurity to capture the phonon mode.

At last, the role of surface corrugations of a particle in the broadening of volume optical phonon modes investigated numerically in present paper is shown to be essentially smaller as compared to volume imperfections. Typically, surface disorder could not even lead to the overlap of the main optical mode, while for separated levels we find $\Gamma_1 \propto \sqrt{c_{imp}}/L^2$ and  $\Gamma_1 \propto \sqrt{c_{imp}} /L^4 $ dependencies for this mode provided that disorder scales with the particle size and does not scale, respectively. The phonon line broadening due surface disorder strongly increases with its quantum number, though.

We observe visually the asymmetry of phonon lines and their non-Lorentzian shapes predicted in paper I.

The paper is organized as follows. In Section \ref{Methods} we formulate the methods we used to study the pure problem and adopt them for numerical treatment of disordered particles. Section \ref{Disorder} sketches sources and peculiarities of disorder in nanopowders and specifies their relations to models considered. Section \ref{WeakR} is devoted to the analysis of weak disorder (including its point-like and smooth versions); it also  addresses the phenomenon of ``mesoscopic smearing''. In Section \ref{StrongR} we investigate strong dilute impurities (including resonant and unitary regimes) and crossover to ``dense'' regime of concentrations. In Section~\ref{Surface} we examine the role of surface corrugations. The last Section~\ref{DiscConcl} contains the summary of our results and their discussion.


\section{Methods}\label{Methods}
In Subsections \ref{DMM-BPM} and \ref{EKFG-S}  we briefly sketch the methods (DMM-BPM and EKFG) we use in order to characterize the Raman scattering. Subsection \ref{Green} is devoted to the description of the Green's functions formalism adapted for disordered particles.

\subsection{Dynamical Matrix Method: Atomistic Approach }\label{DMM-BPM}

Dynamical matrix method (DMM) \cite{born1954dynamical,maradudintheory} provides very good accuracy of derivation of normal vibrational eigenmodes and corresponding eigenfrequencies for multicomponent quantum objects such as molecules and nanoparticles. It consists of solving $3N \times 3N$ matrix eigenvalue problem formulated for the system of $N$ Newtons laws of motion written for $N$ atoms:
\begin{equation}
  \label{DynMat}
 \omega^2 r_{l,\alpha}=\frac{1}{m_l}\sum_{l^\prime=1}^N \sum_{\beta={x,y,z}} \frac{\partial^2 \Phi}{\partial r_{l,\alpha} \partial r_{l^\prime,\beta}}r_{l^\prime,\beta},
\end{equation}
where $r_{l,\alpha}$ is the displacement of the $l$-th atom along direction $\alpha$,  $m_l$ is the mass of the $l$-th atom, and $\Phi$ is the total energy of the particle written as a function of atomic displacements. The masses of atoms could be eliminated from Eqs.~\eqref{DynMat} by the local rescaling of variables
$\xi_l = r_l \sqrt{m_l}$ for varying atomic masses and by the global rescaling $\xi_l = r_l \sqrt{m}$ for identical ones.
%
%
For translationally invariant crystals DMM yields phonon dispersion and polarization. The solution of DMM problem determines displacements $r_{l,\alpha} (n)$ for each eigenmode $n$.

Since we choose nanodiamonds as a testing object for our approach we construct the dynamical matrix utilizing elastic parameters of diamond lattice extracted from the microscopic Keating model \cite{keating1966effect,martin1970elastic,kane1985phonon,steiger2011enhanced}. In this model parameter $\alpha_0$ measures bond rigidity with respect to stretching and parameter $\beta_0$ is responsible for valence angle bending. We use the values $\alpha_0 = 1.068$ Dyn$\cdot$cm$^{-2}$ and $\beta_0 = 0.821$ Dyn$\cdot$cm$^{-2}$ from the Table II of Ref.~\cite{anastassakis1990piezo}.

For the dispersion of optical phonons we utilize conventional expression:
\begin{equation}
\label{eq_dispersion}
\omega_{\bf q}=A+B\cos\,(q  a_0 /2),
\end{equation}
where $a_0=0.357$~nm is the diamond lattice constant.
The Keating model and employed force constants yield  $A+B=\omega_0 \approx 1333 \, \text{cm}^{-1}$ and  $B\approx 85$ cm$^{-1}$. Near the Brillouin zone center dispersion simplifies~\cite{our3}:
\begin{equation}\label{defF}
\omega_{\bf q} = \omega_0 \, [\, 1 - F( q a_0)^2 \, ],
\end{equation}
and $F$ measures spectrum flatness. Comparing Eq.~\eqref{defF} with the Taylor expansion of Eq.~\eqref{eq_dispersion}, we get $F=\frac{B}{8(A+B)}$.


\subsection{Euclidean Klein-Fock-Gordon Equation: Continuous Media Approach}\label{EKFG-S}

Another approach to the Raman spectra of nanoparticles is developed in Ref.~\cite{utesov2018raman}. It utilizes continuous description of long wavelength optical phonons and continuous version of the BPM. Within the framework of EKFG method the phonon wave functions $Y$ satisfy the following differential equation with Dirichlet boundary conditions ($\partial \Omega$ being the particle boundary):
\begin{equation}\label{EKFG}
(\partial^2_t + C_1 \Delta + C_2 ) \, Y = 0, \quad Y|_{\partial \Omega} = 0.
\end{equation}
Eigenfrequencies can be obtained as
\begin{equation}\label{Efreqs1}
  \omega^2 = C_2 - C_1 q^2,
\end{equation}
where $q^2$ is the eigennumber of the problem
\begin{equation}\label{Elaplace}
  \Delta Y + q^2 Y = 0, \quad Y|_{\partial \Omega} = 0.
\end{equation}
Within the range of validity of EFKG approach (small $q \ll a_{0}^{-1}$) the phonon frequency of a mode with generalized quantum number $n$ reads:
\begin{equation}\label{Espec}
  \omega_n \approx \sqrt{C_2} - \frac{C_1}{\sqrt{C_2}} \frac{q^2_n}{2}.
\end{equation}
Since Eq.~\eqref{Espec} should be the quantized version of Eq.~\eqref{defF} one obtains $C_2 = (A+B)^2$ and $C_1 = a_0^2 \, B (A+B)/4$.

The main advantage of EKFG approach is its relative simplicity. Eq.~\eqref{Elaplace} can be solved analytically for cubic, spherical, and cylindrical particles and numerically for other particle shapes using standard methods which contain, e.g., in  MATHEMATICA package~\cite{mathem}.

Importantly, both abovementioned approaches can be easily adopted for numerical treatment. Furthermore, the EKFG approach allows to manipulate with larger particles inaccessible by means of the DMM. In our further analysis we will combine both these methods.

\subsection{Phonon propagator in disordered particles}\label{Green}

Here we adopt the formalism of phonon Green's functions for the treatment of disordered particles.
Let $n$ numerates the eigenstates of a pure particle while the variable $\varepsilon$ spans the energies of eigenstates in the ensemble of disordered particles, the eigenfunctions of these states are $|n\rangle$ and $| \varepsilon \rangle$, respectively. The propagator of the  $n$-th phonon mode has the form  \cite{utesov2014localized}:
\begin{eqnarray}
  D_n(t) &=& i \, \langle \, vac \, | \, \hat{\text{T}} \, ( \, b^{\dagger}_n(t) - b_n(t) \,) \, ( \, b^{\dagger}_n(0) - b_n(0) \, ) \,| \, vac \, \rangle   \nonumber \\ \label{Green1}
  &=& -i \, \theta(t) \, \langle \, n \, | \, e^{-i{\cal H}t} \, | \, n \, \rangle -i \, \theta(-t) \, \langle \, n \, | \, e^{i{\cal H}t} \, | \, n \, \rangle \nonumber  \\  &=& - i \, \sum_\varepsilon | \, \langle \, n \, | \, \varepsilon \, \rangle \, |^2  \, \left[ \, \theta(t) \, e^{-i \varepsilon t} + \theta(-t) \,  e^{i\varepsilon t} \,  \right].
\end{eqnarray}
Here $\hat{\text{T}}$ is the time ordering operator, $b^{\dagger}_{n} (b_{n})$ are the creation (annihilation) operators acting on pure eigenstates, $\theta (t)$ is the Heaviside theta-function, and ${\cal H}$ is the Hamiltonian of disordered problem. After the Fourier transform we obtain the propagator
\begin{equation}\label{Green2}
  D_n(\omega)= \sum_\varepsilon | \, \langle \, n \, | \, \varepsilon \, \rangle \, |^2 \, \left(\frac{1}{\omega - \varepsilon+i 0} - \frac{1}{\omega + \varepsilon - i 0}  \right).
\end{equation}
Near the positive pole its imaginary part reads:
\begin{equation}\label{ImGreen1}
 \text{Im} \, D_n(\omega) = -\pi \, \sum_\varepsilon \, \delta \, (\omega- \varepsilon) \, | \, \langle \, n \, | \, \varepsilon \, \rangle \, |^2.
\end{equation}
Averaging Eq.~\eqref{ImGreen1} over disorder configurations (which is denoted by the overline) we calculate the density of states $\rho \, (\omega)$:
\begin{equation}\label{ImGreen2}
 \overline{\text{Im} \, D_n(\omega)}  = -\pi \, \rho \,(\omega) \, \overline{ | \, \langle \, n \, | \, \omega \, \rangle \, |^2}.
\end{equation}
This quantity multiplied by $-\pi^{-1}$ represents the spectral weight of the broadened vibrational eigenmode. Using the Lorentzian approximation for the spectral weight we obtain the linewidth for the $n$-th mode.

In order to perform disorder averaging in Eq.~\eqref{ImGreen1} one should specify the type of disorder. Within the DMM approach we deal with eigenfunctions $r(k,n)_{l,\alpha}$, where $k$ stands for the index of disorder realization (we reserve the value $k=0$ for pure particle). Then Eq.~\eqref{ImGreen2} becomes:
\begin{eqnarray}\label{ImGreenDMM}
  \overline{\text{Im} \, D_n(\omega)} &=& - \frac{\pi}{N_{c}} \, \sum^{N_{c}}_{k=1} \, \sum^{3N}_{n^\prime=1} \, \delta \, (\omega -\omega_{n^\prime})  \\ &\times &  \left[ \, \sum^N_{l=1} \, \sum_{\alpha=x,y,z} \, r(0,n)_{l,\alpha} \, r(k,n^\prime)_{l,\alpha} \, \right]^2, \nonumber
\end{eqnarray}
where $N_c$ is the number of configurations. In order to distinguish visually between DMM-BPM and EKFG formulas we denote the eigenfunctions as  $r(k,n)_{l,\alpha}$ in the former and as $Y_{n}(k,\mathbf{r})$ in the latter case which differs from notations of Ref.~\cite{our3}.
In real calculations we replaced the upper limit in the inner sum (the number of eigenmodes $3N$) by the energy cutoff $\omega_c = 1000$ cm$^{-1}$ in order to prevent from the mixing of optical and acoustic modes at large momenta. It is justified by the fact that we never observed the broadening reaching the value $\sim 10^2$ cm$^{-1}$  even for unrealistically strong disorder.

For continuous EKFG calculations the analog of Eq.~\eqref{ImGreenDMM} reads:
\begin{eqnarray}\label{ImGreenEKFG}
  \overline{\text{Im} \, D_n(\omega)} &=& - \frac{\pi}{N_{c}} \, \sum^{N_{c}}_{k=1} \, \sum_{n^\prime=1}^{N^\prime} \, \delta \,(\omega -\omega_{n^\prime}) \\ &\times&  \left[ \, \int_\Omega d^3 \mathbf{r} \, Y_n(0,\mathbf{r}) \, Y_{n^\prime}(k,\mathbf{r}) \, \right]^2, \nonumber
\end{eqnarray}
where $\Omega$ is the volume of a particle.
The upper limit $N^\prime$ of the inner sum in Eq.~\eqref{ImGreenEKFG} is chosen from the condition for characteristic ``wave vector'' $k_c =\frac{{\rm max} Y'}{{\rm max}Y}$ to be smaller than the size of the Brillouin zone $\pi/a_0$, where the prime stands for spatial derivative.

In Section~\ref{Methods} we briefly describe discrete DMM and continuous EKFG methods of evaluation of vibrational eigenmodes in pure particles  we intend to utilize for the treatment of disordered problem. We also express parameters appearing in these methods via the parameters of microscopic Keating model. Furthermore, we adopt the Green's functions formalism for numerical treatment of optical phonons in disordered finite particles, and connect it with both abovementioned methods.


\section{Disorder}\label{Disorder}\label{Disorder}

In Subsection~\ref{Preliminary}  we present a qualitative picture of various types                                                                                                                                 of disorder existing in nanoparticles. The peculiarities of their numerical investigation within the framework of approaches outlined in Section~\ref{Methods} are discussed in Subsections \ref{BinaryD}, \ref{GaussianD}, \ref{EKFG-D}, and \ref{SurfaceD}.

\subsection{Preliminary Remarks}
\label{Preliminary}

\begin{figure}
  \centering
  \includegraphics[width=6.4cm]{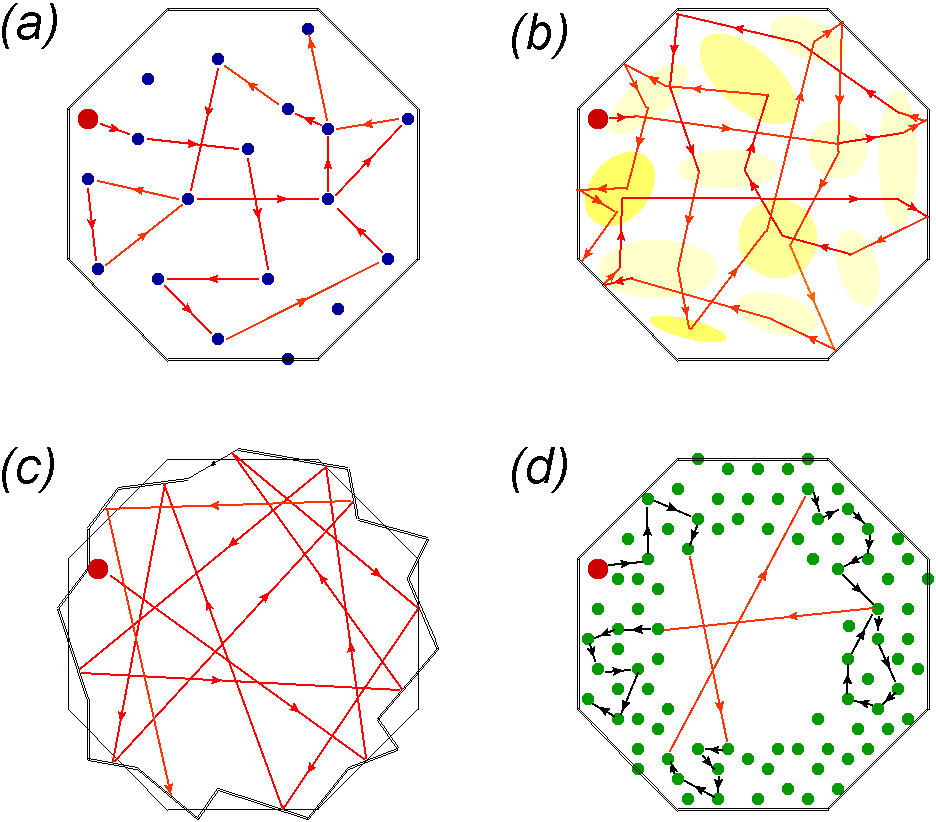}
  \caption{Sketch of four types of disorder contributing to phonon damping in nanoparticles.
  (a) Point-like impurities located in the interior of a particle.
  Phonons propagate through the particle and sometimes scatter off the impurity.
  If the scattering occurs frequently (as in this picture), the phonon levels overlap.
  (b) Smooth disorder in the volume of a particle. The characteristic disorder scale $\sigma$ is shorter than the particle size $L$.
  (c) Surface corrugations which can be considered as random faceting of a particle. They lead to chaotization of propagating waves.
  (d) The shell of amorphous phase near the surface of a particle. Vibrations propagating within this shell are not the phonons.
	\label{defects}}
\end{figure}

When addressing phenomena responsible for disorder induced broadening of the Raman peak in nanoparticles one should mention four principal mechanisms.
The first one is due to ``ordinary impurities''. Substitutional impurities in diamonds are realized as distortions of the dynamical matrix stemming from random replacements of carbon atoms  by another isotope of carbon (weak point-like impurities) or by another sort of atoms (e.g., by nitrogen which forms so-called NV centers~\cite{doherty2013nitrogen} identified below as strong point-like impurities). These local variations of masses and rigidities scatter vibrational modes and therefore provide their damping (see Fig~\ref{defects}a). Although physically masses and rigidities vary simultaneously, in Ref.~\cite{our3} for the sake of simplicity only the masses have been taken to be random.

The second mechanism is related to large-scale inclusions (gaseous or solid) often existing in nanoparticles, lattice distortions (dislocation, etc.), and slow changes of characteristics of the crystal caused by evolution of external parameters (pressure, temperature, chemical composition of the atmosphere, etc.) during the time of crystal growth. All these influences have the large-scale structure and could be regarded as a ``colored noise'' with certain spatial scale $\sigma$ exceeding the interatomic distance $a_0$ and comparable with the particle size $L$ (see Fig~\ref{defects}b). We describe this type of disorder qualitatively by means of a smooth random (Gaussian) potential.

The third way to incorporate disorder into the problem referred to as ``surface corrugations'' is specific for finite particles. It arises if one introduce the irregularities of the particle surface, the phonon scattering within the particle is due to surface reflections only whereas the interior of the crystal may be perfect for wave propagation  (see Fig~\ref{defects}c). Surface scatterings broaden the phonon line, the details of broadening depend on the character of surface roughnesses. The correspondence between the propagation of electromagnetic (or electron) waves in irregular cavities and in regular cavities with spatial disorder has been widely discussed in late 90s in terms of chaotization in (both quantum and classical) ``billiards''~\cite{Billiards2010}.

The fourth mechanism which leads to the Raman peak broadening in nanoparticles originates from surface amorphization (see Fig~\ref{defects}d)  well-documented for nanodiamonds \cite{ferrari2004raman,popov2013features,dideikin2017rehybridization,koniakhin2020evidence}. It takes place within the near-surface shell of a crystal due to crystal interference with surrounding media during its growth and/or aging. The amorphous surface shell is known to exist in bulk crystals,  either, but its importance increases with increasing the surface-to-bulk ratio, i.e., in nanoparticles. This mechanism implies very strong disorder in the near-surface shell where even the notion of propagating wave (phonon) modes looses its meaning due to the lack of translational invariance. It is the only mechanism we shall not address in this paper.


We believe that in reality there realizes certain combination of all four aforementioned mechanisms.

\subsection{Binary Disorder in Discrete Model}\label{BinaryD}

It is important to define a  clear and meaningful measure of disorder strength. Consider first the simplest binary disorder, when the lattice sites are occupied  with relative probability $1- c_{imp}$ by atoms with masses $m$ and  with relative probability $c_{imp} $ by randomly distributed impurities with masses $m + \delta m$. Here $c_{imp}=N_i /N$ is the dimensionless concentration, $N_i$ is the total number of impurities and $N$ is the total number of lattice sites in a particle. Then at low concentration $c_{imp} \ll 1$ and small mass difference $\delta m / m \ll 1$ the disorder strength parameter could be introduced as follows:
\begin{equation} \label{DisW}
    S = c_{imp} \left( \frac{\delta m}{m} \right)^2.
\end{equation}
The mean variation of binary disorder is not equal to zero, $\langle \delta m_i \rangle \neq 0$.
The strength of strong impurities with $|\delta m| \gtrsim m$ should be measured by the parameter
\begin{equation}\label{DisS}
U= \frac{\delta m}{m + \delta m},
\end{equation}
still, under the condition $c_{imp} \ll 1$. If $|U| \sim 1$ {\it and} $c_{imp} \sim 1$, than the notion of translational invariance (in any sense) disappears, and the solid becomes amorphous.

\begin{figure}[h]\label{fig4peaks}
\centering
  \includegraphics[width=0.90\linewidth]{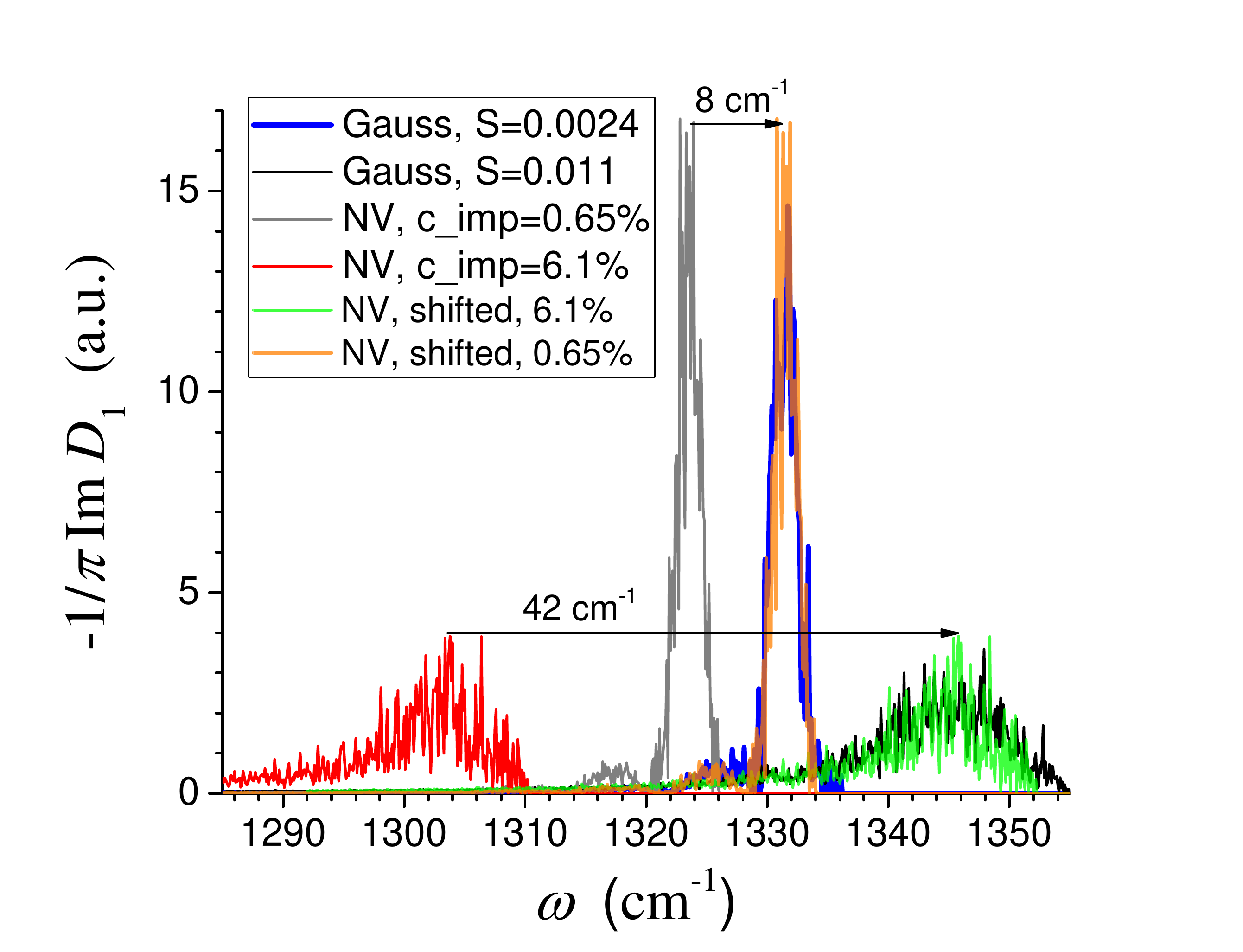}
  \caption{\label{fig4peaks} The spectral weight of the first vibrational mode obtained using the DMM method for Gaussian disorder (blue and black curves) and for binary disorder in the form of NV centers (gray and red curves) drawn for two values of disorder corresponding to separated and overlapped regimes. Binary curves are significantly shifted with respect to the Gaussian ones due to nonzero $\langle \delta m_l \rangle$; however, after the substraction of these shifts (orange and green curves) Gaussian and binary line shapes became almost indistinguishable. Also, the line shapes calculated for the overlapped regime are asymmetric in agreement with the theory of paper I}
\end{figure}

For diamonds, the binary distribution appears in two variants: as the isotopic-induced disorder and in the form of substitutional nitrogen impurities. In the former case $m$ corresponds to $^{12}$C atom  and $m + \delta m$ is the $^{13}$C atomic mass which yields $\delta m/ m=(13-12)/12=1/12$. When the nitrogen atoms are considered, the effect is not so obvious. Normally, nitrogen is accompanied in diamonds by the vacancy in carbon lattice (NV center). These NV centers in nanodiamond are in fact a molecule-like complexes embedded into the carbon atoms surrounding and having the rich characteristic vibrational level structure \cite{doherty2013nitrogen,kehayias2013infrared}. The nitrogen atom possesses 5 electrons with the spatial structure of sp$^3$ hybridization in the outer shell: 3 of them form covalent bonds with neighboring carbon atoms and 2 electrons related to the vacancy constitute a pair (accompanied by an outer electron to form NV$^-$ center). The spectra of NV centers are quite intricate and to access the fine structure it should be studied by \textit{ab initio} methods \cite{gali2011ab,ekimov2019ab,gali2019ab}. As far as the long wavelength optical phonons are considered, all these peculiarities do not play significant role, and phonons feel NV defects as point-like disturbances in the dynamical matrix of a diamond nanoparticle. Within the DMM approach we model NV centers as a vacancy neighboring the impurity with $\delta m/m = 0.17$; however, even simpler presentation via binary disorder with unitary defects $U \rightarrow -\infty$ yields a good result (see below).

\subsection{Gaussian Disorder in Discrete Model}\label{GaussianD}

For a Gaussian disorder, the masses of atoms are randomly distributed according to the Gaussian law around some mean value. Although precisely this form of disorder is rarely realized in practice, the model is very popular among the theorists for its analytical convenience. Moreover, the solution of Gaussian problem reveals the majority of generic features shared by many physical types of disorder (see, e.g., Fig.~\ref{fig4peaks}). Therefore, the numerical studying of Gaussian disorder is the most straightforward way to verify the predictions of theory simultaneously staying on physical grounds.

The definition Eq.~\eqref{DisW} can be generalized onto arbitrary distribution of impurity masses:
\begin{equation}\label{delta m 2 av}
S = \frac{ \langle (\delta m_l)^2 \rangle }{m^{2}},
\end{equation}
where nonzero contributions come from impurity sites only making the quantity $S$ proportional to the impurity concentration $c_{imp}$.
The distribution function for a Gaussian disorder with $|\delta m|/m \ll 1$  is given by:
\begin{equation} \label{distrib}
   {\cal F} \, (\delta m_l) = \frac{1}{\sqrt{2 \pi S}} \exp{\left[-\frac{(\delta m_l/ m  )^2}{2 S}\right]}.
\end{equation}
For the purposes of DMM-BPM method we model the weak Gaussian disorder as follows. The carbon-normalized inverse impurity mass $\tilde{m}^{-1}$ of $l$-th atom is chosen to be randomly varying within the interval  $[ \, 0.25, \, 1.75 \, ]$ with the probability
\begin{equation}
    {\cal F} \, (\tilde{m}_l^{-1}) \propto \exp\left(- \frac{(\tilde{m}_l^{-1}-1)^2}{2 S} \right).
\end{equation}
This procedure does not deliver zero for the average mass variation. Therefore, we add to each mass the quantity
\begin{equation} \label{dissubtr}
    \Delta \tilde{m}^{-1} = 1 - \frac{1}{N}\sum_{l=1}^N \tilde{m}_l^{-1},
\end{equation}
thus providing $\langle \delta m_{l} \rangle = 0 $.

\subsection{Disordered continuous EKFG Model}\label{EKFG-D}

Disorder can be incorporated into continuous model by several means. The {\it point-like} impurities of discrete DMM-BPM approach are mapped onto the spatial variation of parameters $C_{1,2}$ in EKFG equation. However, since the term with $C_1$ in Eq.~\eqref{EKFG} possesses additional smallness due to spatial derivatives  we can vary only $C_2$:
\begin{equation}\label{Edisorder}
  \delta C_2 (\mathbf{r}) = - \frac{\delta m (\mathbf{r})}{m + \delta m (\mathbf{r})} C_2.
\end{equation}
Here $m$ and $C_2$ stand for their values in a pure particle. Then for Born impurities one has:
\begin{equation}\label{Edisweak}
  \langle \delta m (\mathbf{r}) \rangle =0, \quad \frac{\langle \delta m (\mathbf{r}) \delta m (\mathbf{r}^\prime) \rangle}{m^2} = S \, V_0 \, \delta(\mathbf{r}^\prime-\mathbf{r}),
\end{equation}
where $V_0$ is the unit cell volume and $S$ is the dimensionless strength of impurities given by Eq.~\eqref{delta m 2 av}.

In our numerics we use MATHEMATICA package~\cite{mathem}. For Born impurities we reformulate the definition of momentum $q$ in Eq.~\eqref{Elaplace} which can be rewritten in the form:
\begin{equation}\label{Elaplace2}
    \Delta Y + \frac{\delta \, C_2(\mathbf{r})}{C_1} \, Y + q^2 \, Y = 0, \quad Y|_{\partial \Omega} = 0.
\end{equation}
Notice that the eigenfrequencies still can be found from Eq.~\eqref{Efreqs1}, where $C_2$ corresponds to the pure particle. Then the procedure of generating new disorder configurations and numerical solving Eq.~\eqref{Elaplace2} should be performed repeatedly. Obtained eigenfunctions and eigenfrequencies are utilized in subsequent calculation of the disorder-induced phonon line broadening (see Subsection~\ref{Green}).

\begin{figure}[t]
\includegraphics[width=0.44 \linewidth]{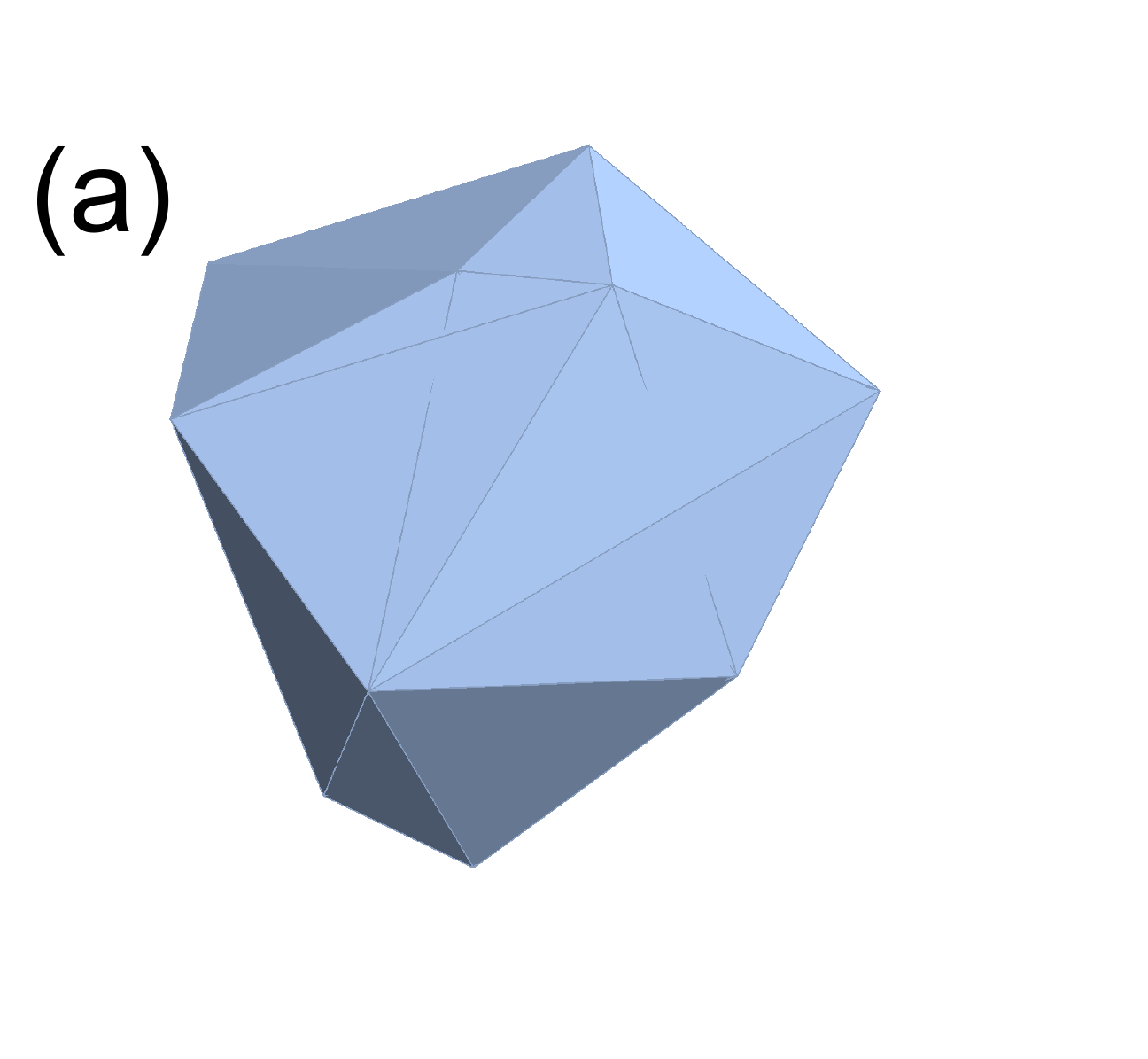}\includegraphics[width=0.36 \linewidth]{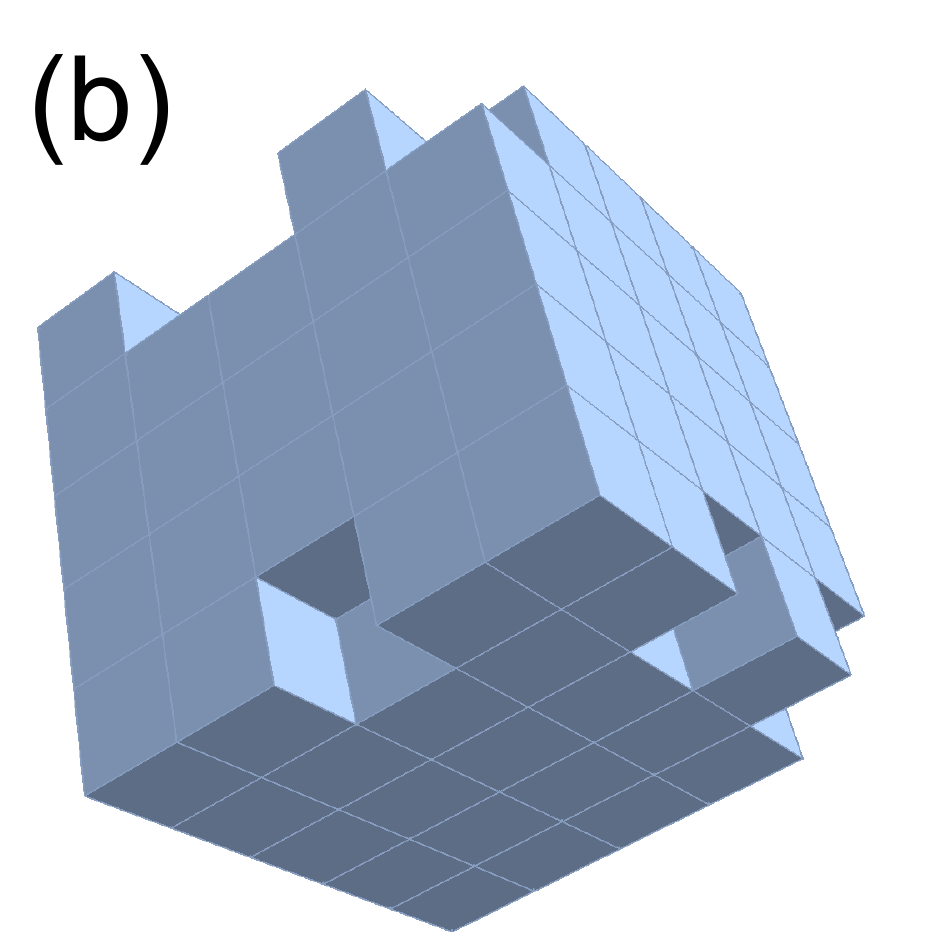}
\caption{\label{SurfCorr} (a) The particle with the random surface made of triangular facets (``peeled apples'' model). This object is an example of the convex irregular polyhedron. (b) The particle made of cubic bricks with randomly removed bricks on the surface (``nibbled apples'' model). }
\end{figure}

In order to adopt EKFG approach for the treatment of a {\it smooth} disorder let us introduce $N_{def}$ defects providing the random Gaussian potential in the form:
\begin{equation}\label{Smooth1}
  \delta C_2(\mathbf{r}) = \sum_{i=1}^{N_{def}} \frac{ d_i \, C_2 }{(2\pi \, \sigma^2)^{3/2}} \,\, e^{-\displaystyle{\frac{(\mathbf{r}-\mathbf{r}_i)^2}{2 \sigma^2}}},
\end{equation}
where $\mathbf{r}_i$ are the centers of uncorrelated impurity potentials and $d_i$ are corresponding (Gaussian distributed) strength constants  obeying the following conditions:
\begin{equation}\label{Smooth2}
  \langle \, d_i \, \rangle = 0, \quad \langle \, d_i d_j \, \rangle = \delta_{ij} \, S.
\end{equation}
One can check that $\delta C_2 (\mathbf{r}) \, \delta C_2 (\mathbf{r}^\prime) \propto S$, and, performing disorder averaging $\langle \delta C_2 (\mathbf{r}) \delta C_2 (\mathbf{r}^\prime) \rangle$, the standard deviation is equal to $\sigma$, the latter result is valid if one neglects the boundary effects.
Thus, we arrive to the problem investigated theoretically in paper I.

When applying continuous EKFG approach in order to investigate more involved case of rare {\it strong} impurities we introduce $N_{imp}$ unit cells in a particle ($N_{imp} \ll N$)   with fixed large value of $|\delta m|$ inside every cell.

\subsection{Surface corrugations}\label{SurfaceD}

\begin{figure}[h]
\centering
  \includegraphics[width=0.96\linewidth]{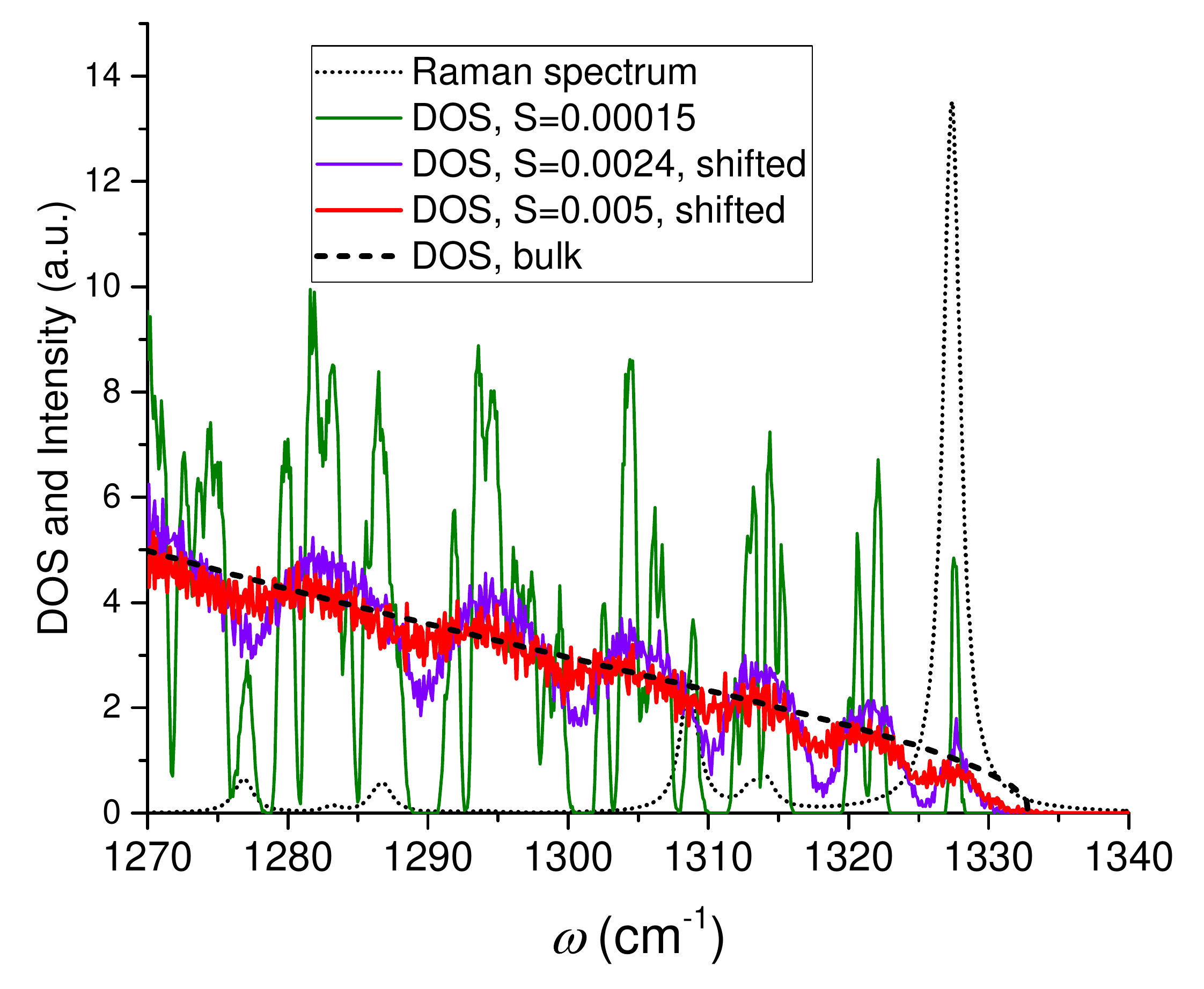}
  \caption{\label{fig_DOS} Phonon density of states vs frequency. The bulk pure DOS is shown by dashed black curve. Disordered DOS are calculated for 3~nm spherical diamond particles with the use of DMM approach for weak Gaussian disorder. At small $S=0.00015$ the DOS has a comb-like structure, and resulting Raman peaks are shown by dotted black curve. Crossover between the regimes of separated and overlapped levels occurs between $S=0.0024$ and $S=0.005$.}
\end{figure}

We adjust canonical EKFG approach for the analysis of influence of particle surface irregularities on the phonon linewidth by solving the Laplace eigenproblem with Dirichlet boundary conditions given by Eq.~\eqref{Elaplace} for the shape of a boundary $\partial \Omega$  randomly varying from particle to particle under the constraint to preserve the particle volume. We examine in details two particular models of surface corrugations. The first one is the random triangular faceting of a cubic particle which yields the convex irregular polyhedron (hereinafter, the ``peeled apples'' model, see Fig.~\ref{SurfCorr}a). For the second model we construct the particle using cubic bricks of certain size, some bricks on surface are randomly removed with probability $c_{imp}$ (``nibbled apples'' model, see Fig.~\ref{SurfCorr}b). These models differ by the  type of surface irregularities whereas the volume of a particle supposed to be clean in both cases.

To summarize this Section, we list four types of disorder physically realized in nanoparticles; three of them are examined in present paper. We argue that the isotopic disorder should be attributed as weak point-like impurities and the NV centers (nitrogen + vacancy) as strong point-like ones. We specify how three types of disorder mentioned will be modelled in our numerical calculations.


\section{Results: Weak Impurities}\label{WeakR}

In this Section we present the results of numerical modelling for weak (both point-like and smooth) disorder and
compare these results with theoretical predictions made in paper I.  We also discuss the phenomenon of ``mesoscopic smearing''  not addressed in paper I.  The disorder assumed to be weak if not only the condition $S \ll 1$ is fulfilled but also its constituents $c_{imp}$ and $(\delta m / m)^2$ are much smaller than unity {\it independently}.

\subsection{Density of states. Spectral Weight}\label{DOS}

Let us discuss the behavior of the phonon density of states (DOS) in disordered nanodiamonds. The bulk DOS reveals the van Hove singularity at $\omega \to \omega_0$:
\begin{equation}
    \label{eq_DOS_analyt}
    \rho (\omega \to \omega_0) = \frac{ \theta (\omega_0 - \omega) \, \sqrt{\omega_0 - \omega}}{4 \pi^2 \, (F \omega_0)^{3/2}}.
\end{equation}
\begin{figure}[t]
\centering
  \includegraphics[width=0.90\linewidth]{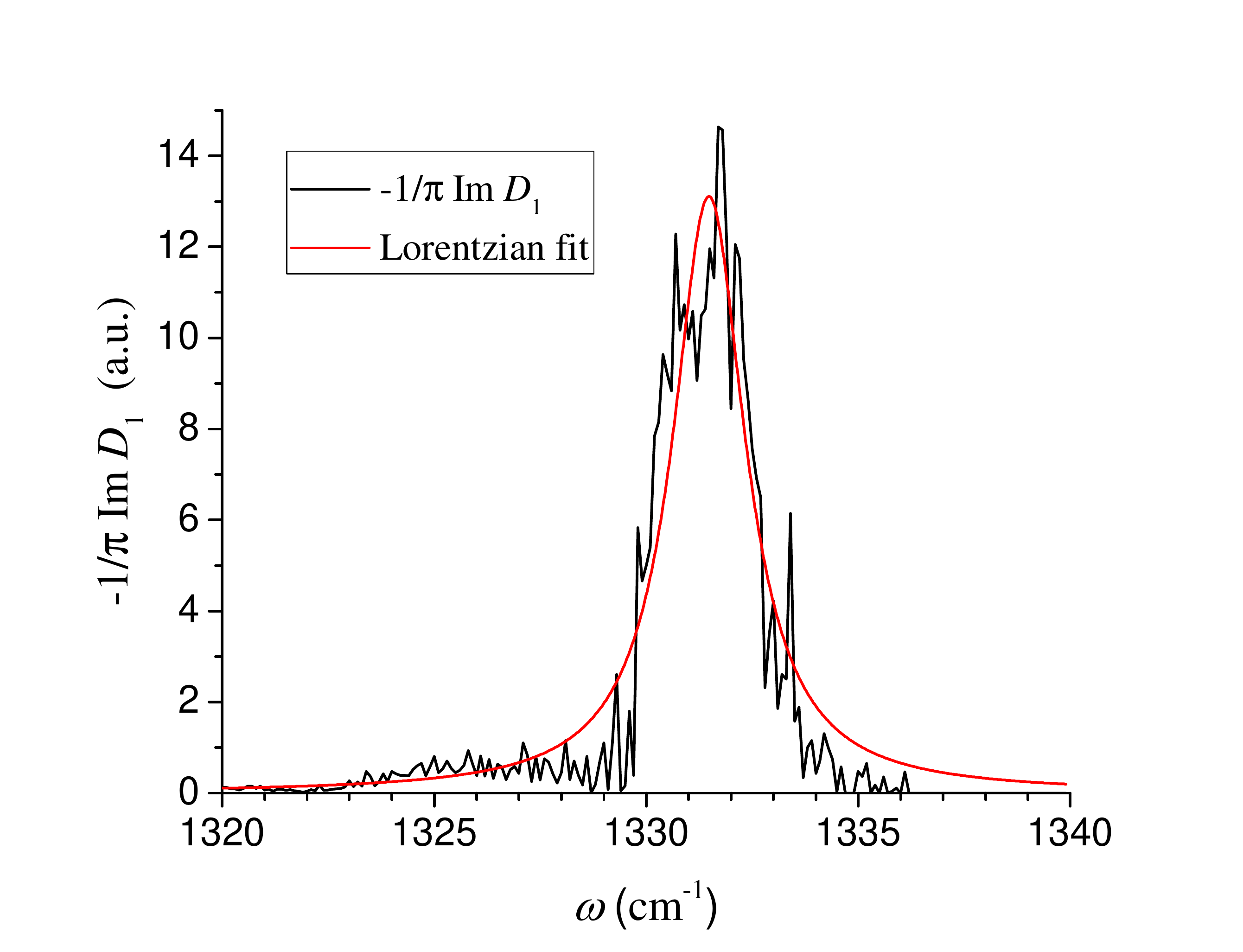}
  \caption{\label{fig_fit} Detailed structure of spectral weight for the first phonon eigenmode $D_1$ in disordered nanoparticle. Disorder strength $S=0.0024$ and other parameters are the same as in Fig.~\ref{fig_DOS}, where this eigenmode is seen as tiny blue hump under the Raman peak.  Although this value of $S$ is close to the crossover one, the shape of a peak still resembles semi-circle more than Lorentzian (cf. Fig.~5 of paper I).}
\end{figure}
More involved formula applicable in wider frequency range could be found in Ref.~\cite{koniakhin2018raman}. The bulk DOS together with DOS functions calculated numerically for 3~nm disordered spherical diamonds are plotted in Fig.~\ref{fig_DOS} as functions of frequency for different values of disorder, the disorder-induced shift of $\omega_0$ is subtracted by hands. At smallest $S$ the phonon lines acquire just a little broadening much smaller than energy spacings between different states. The resulting DOS has a comb-like structure similar to the DOS of a single particle. This structure of DOS reproduces itself in the Raman spectrum, as it occurs, e.g., for fullerenes \cite{bethune1991vibrational}. Observation of a comb-like Raman spectrum for one nanoparticle or ensemble of very small, clean, and equal-sized nanoparticles is very challenging but intriguing experimental task.

 When the disorder strength grows up the phonon lines start to overlap (see Fig.~\ref{fig_DOS}). However, every line can  be
 resolved and the first mode remains well separated. Qualitative change in DOS visible by eye shows up between $S=0.00024$ and $S=0.005$. The first mode starts to overlap with its neighbor  whereas other lines are strongly overlapped and smeared out into the bulk-like DOS with some small features on the top. For stronger disorder the DOS in nanoparticle is almost the same as in the bulk diamond but it has pronounced tail at $\omega > \omega_0$ hiding the van Hove singularity located at $\omega_0 \approx 1333$~cm$^{-1}$.

\begin{figure}[h]
\centering
  \includegraphics[width=0.90\linewidth]{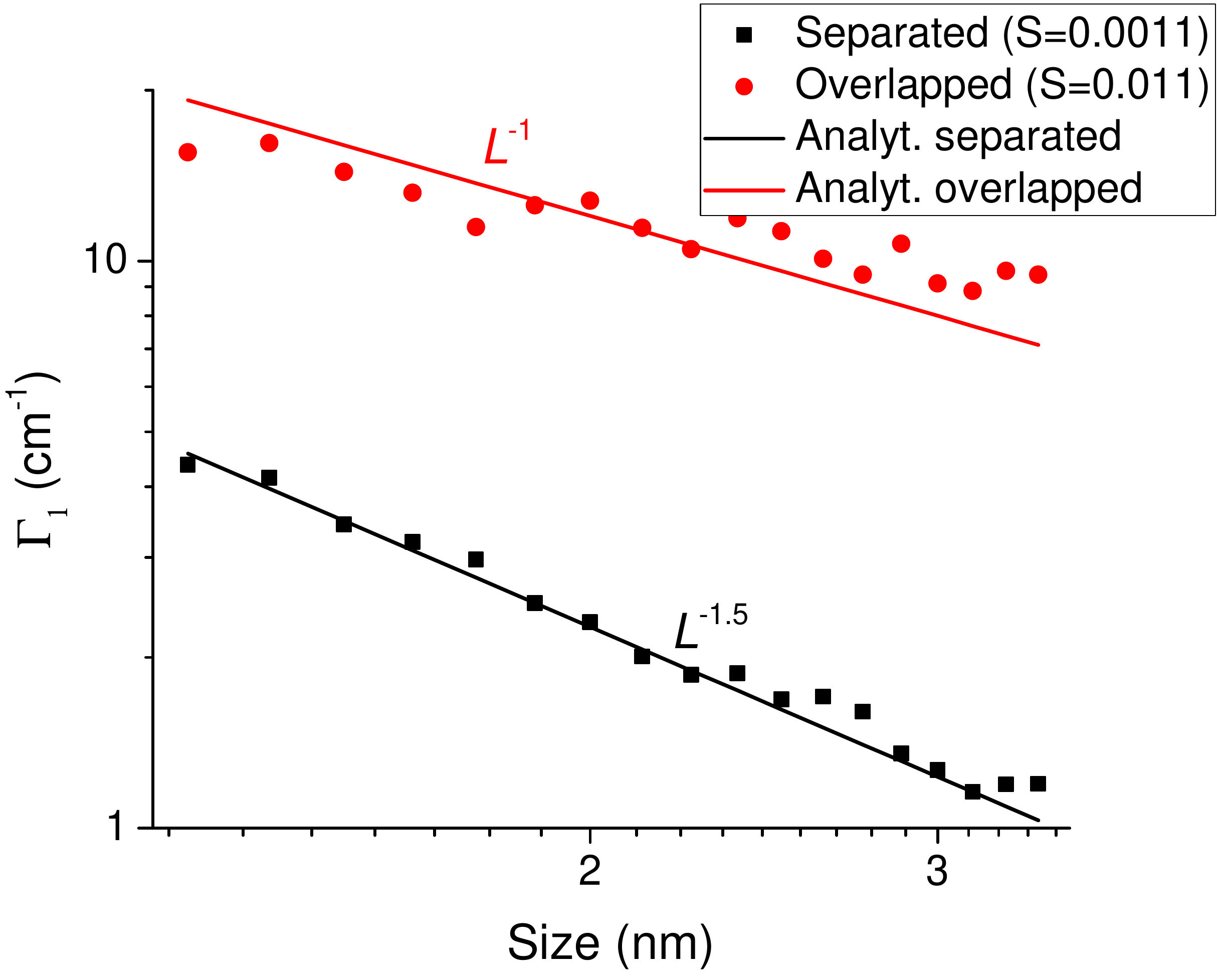}\label{figure}
  \caption{\label{fig_gamma_on_size} The broadening parameter for the first optical phonon eigenmode $\Gamma_1$ as a function of particle size $L$ obtained within the DMM-BPM scheme for spherical diamond particles subject to weak Gaussian disorder. Black squares and red dots correspond to disorder strengths $S=0.0011$ and $S=0.011$, respectively. The range of $L$ presented in Figure covers the regime of separated levels in the former case and the regime of overlapped levels in the latter one, the lines depict theoretical predictions for these two regimes.}
\end{figure}

The picture described above for disordered DOS in nanoparticles is in one-to-one correspondence with the picture of Raman peak structure in nanoparticles presented in paper I. Indeed, the discrete vibrational eigenmodes (see, e.g., Fig.~\ref{fig_fit}) which we call (not entirely accurate) ``phonons'' constitute both the disordered DOS in nanoparticles and the Raman peak (peaks). The only difference is that in the latter case some of these lines are suppressed due to symmetry properties of related eigenfunctions, and the rest of them are re-weighted with matrix elements of the photon-phonon interaction. These eigenmodes can exist either in separated or in overlapped regime depending on disorder strength and particle size. Moreover, since the distance from the first triple-degenerate mode to its closest neighbor is well above any other inter-level spacing in the spectrum the particle can exist in mixed state (cf. Fig.~\ref{fig_DOS}, magenta curve), which provides a {\it wide} crossover between purely separated and completely overlapped regimes.

\subsection{Linewidth. Crossover Scales}

Here we present a comparison between the results of analytical calculations of paper I and the numerics of the present paper concerning the phonon line broadening.

Analytical formulas for the phonon linewidths reads
\begin{equation} \label{GWsepar}
   \Gamma_n =  \omega_n \, \mu_n (p) \, \sqrt{S} \, \left( \frac{a_0}{L} \right)^{3/2},
\end{equation}
for separated levels and
\begin{equation}\label{GWover}
\Gamma_n = \omega_{n} \, \nu_n(p) \, S \, \frac{a_0}{L}.
\end{equation}
for overlapped ones. Here $\mu_n (p)$ is shape $p$ and quantum number $n$ dependent coefficient defined by
\begin{equation}\label{mu}
\mu_n^2 (p) = P^3_p \, \frac{ \, N}{128} \, \sum_{l,\alpha} \left( \, r(n)_{l,\alpha} \, r(n)_{l,\alpha} \, \right)^2,
\end{equation}
\begin{figure}[t]
\centering
  \includegraphics[width=0.80\linewidth]{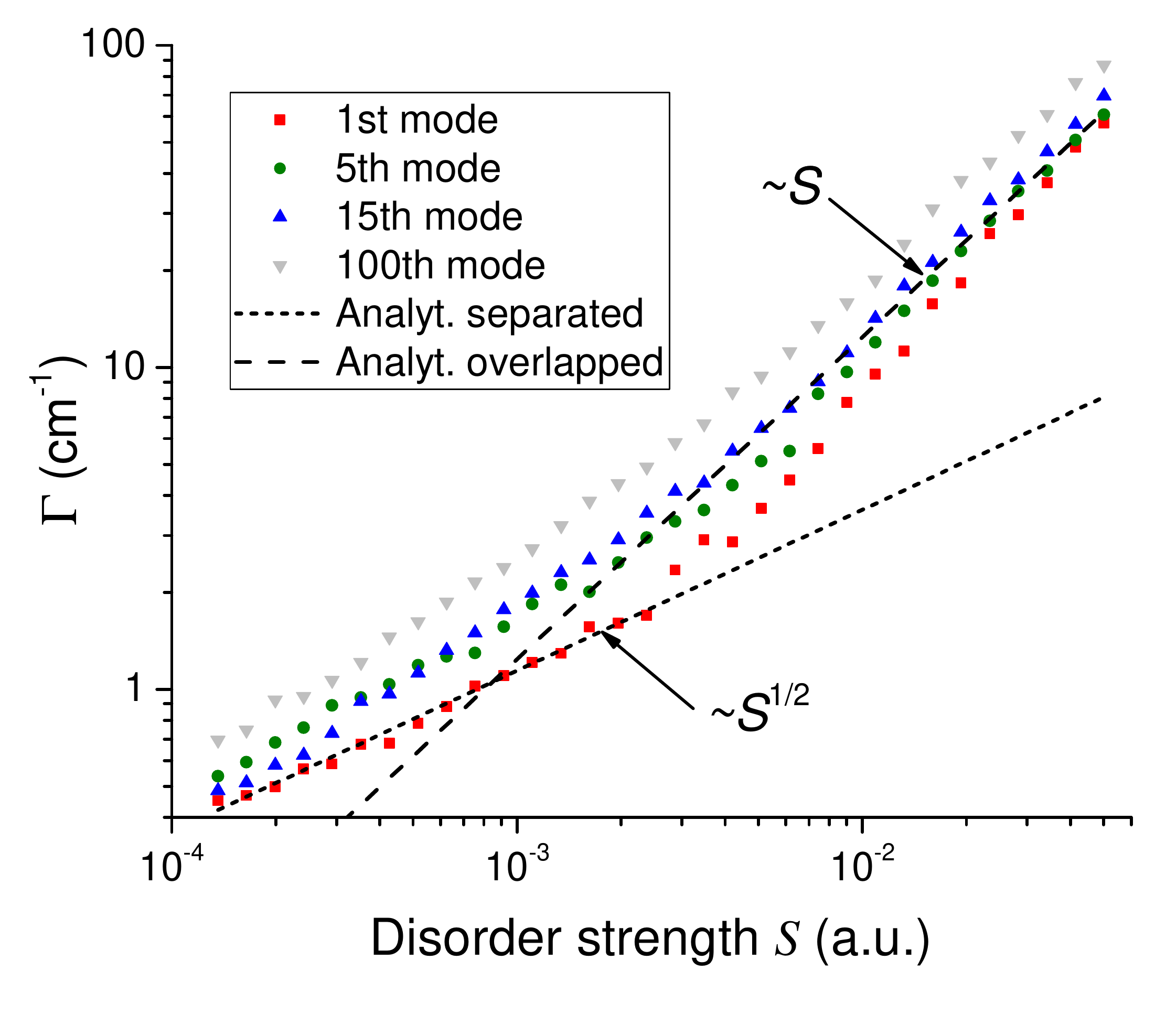}\label{figure}
  \caption{\label{fig_weak_gauss} The broadening parameter $\Gamma_n$ for several optical phonon eigenmodes  as a function of disorder strength $S$ obtained within the DMM approach for 3~nm spherical diamonds. At small $S$ the regime of separated levels takes place (dotted line), where $\Gamma_n \propto \sqrt{S}$  and reveals a weak quantum number dependence, whereas at larger $S$ the linewidth follows the asymptote $\Gamma_n \propto S$ (dashed  line), typical for overlapped levels. Crossover between these regimes occurs at $S \sim 0.005$.}
\end{figure}
where $P_p$ converts the linear size of a particle with $p$ facets into the diameter of a sphere $L$ containing the same amount of atoms, the sum in Eq.~\eqref{GWsepar} runs over all atoms in a particle. Furthermore, $\nu_n (p) \propto 1/ \, 64F$ and strongly depends on the quantum number $n$.
The width of the Raman peak inversely proportional to the particle size has been extracted from experimental data in Ref.~\cite{yoshikawa1995raman}, where the particles with $L \sim 10^2$~nm have been analyzed.

First, we test the linewidth dependence on the particle size $L$ predicted by Eqs.~\eqref{GWsepar} and \eqref{GWover}. Using DMM approach and Gaussian distribution of disorder we investigated numerically the broadening of the first phonon line in spherical diamond particles as a function of particle size for two values of disorder strength supposedly corresponding to regimes of separated ($S=0.0011$) and overlapped ($S=0.011$) levels. The results are plotted in Fig.~\ref{fig_gamma_on_size}. We observe predicted power-law dependencies $\Gamma_1 \propto L^{-3/2}$ and  $\Gamma_1 \propto L^{-1}$ for these two cases, respectively. Notice that not only
the functional dependence of $\Gamma_1$ but also the numerical prefactors are in good agreement with the theory of paper I.

Second, in order to examine disorder strength and quantum number dependencies in Eqs.~\eqref{GWsepar} and \eqref{GWover}, we study numerically the broadening parameters $\Gamma_n$ for several phonon modes with different quantum numbers versus disorder strength parameter $S$ in spherical 3~nm diamond particles, the method used was the same as for Fig.~\ref{fig_gamma_on_size}. The result is plotted in Fig.~\ref{fig_weak_gauss}. We mention very good agreement between the numerics and the theory for the first phonon linewidth $\Gamma_1$ including functional dependencies and crossover scale. For higher modes transition between the regimes occurs smoother than for the first one manifesting wide crossover area and (for highest mode) overlapped regime for all $S$ considered. The tendency for higher modes to have larger linewidths in the regime of overlapped levels predicted in paper I is correctly reproduced in our numerical experiment; however, we observed that the character of this growth is overestimated by analytical theory. Nevertheless, the linewidths $\Gamma_n$ growing with increasing of their quantum numbers $n$ is an important ingredient of our approach. It has been demonstrated in Ref.~\cite{ourShort} where incorporating of this phenomenon essentially improved the $\chi^2$ criterion as compared to the fit with $n$-independent linewidths.

Notice that for the fit of $\Gamma_1$ on separated levels we used the value $\mu_1 \approx 0.33$ in Eq.~\eqref{GWsepar} which is nearly
1.5 times smaller than its value extracted from exact DMM eigenfunctions. This discrepancy we attribute to the Lorentzian approximation we used in analytics, whereas the semi-circle form (see paper I) gives additional $\sqrt{3}$ factor which solves the problem. Moreover, it is seen in Fig.~\ref{fig_fit} that the real phonon spectral weight is not just non-Lorentzian but even asymmetric. We believe, that the r.h.s. of  Eq.~\eqref{GWsepar} should be multiplied by the factor $\sqrt{3}$ caused by imperfection of Lorentzian approximation.

The third issue we would like to discuss in this Subsection concerns crossover scales between the regimes of  phonon line broadening. The estimates for these scales presented in paper I for the mean particle size $L$
\begin{equation}\label{CrosL}
{\cal L}_c \sim \frac{a_0}{S}
\end{equation}
at given disorder strength and for disorder strength $S$
\begin{equation} \label{CrosS}
{\cal    S}_{c} \sim \frac{a_0}{L}
\end{equation}
at fixed particle size yield just qualitative understanding of this issue without answering the question: is this particular phonon mode separated from others or overlapped with them, the answer depends also on shape and quantum number dependent prefactors omitted in Eqs.~\eqref{CrosL} and \eqref{CrosS}.
To give some feeling of numbers, we rewrite here expression for ${\cal L}_c$ obtained in the Appendix of paper I for the first vibrational mode of a cubic particle which includes all numerical and parametric prefactors:
\begin{equation}\label{Scale1}
  {\cal L}_{c} =  \,15  \pi^4 F^2 \, \frac{a_0}{S}.
\end{equation}
For a diamond, the flatness parameter $F \approx 0.008$, and the prefactor in front of the model-free ratio in Eq.~\eqref{Scale1} is about 0.094. It means that for reasonable amount of disorder, say, between $S=0.001$ and $S=0.05$ the crossover particle size ${\cal L}_c$ varies from 1.9$a_0$ to 94$a_0$, or, in metric units, from 0.67~nm to 34~nm. It is precisely the range of parameters intensively studied in recent experiments.

\subsection{Smooth Disorder}\label{Smooth}

In this Subsection we discuss the effect of a smooth weak disorder characterized by the length scale $\sigma$ on optical vibrations in nanoparticles.

\begin{figure}[h]
\centering
  \includegraphics[width=0.80\linewidth]{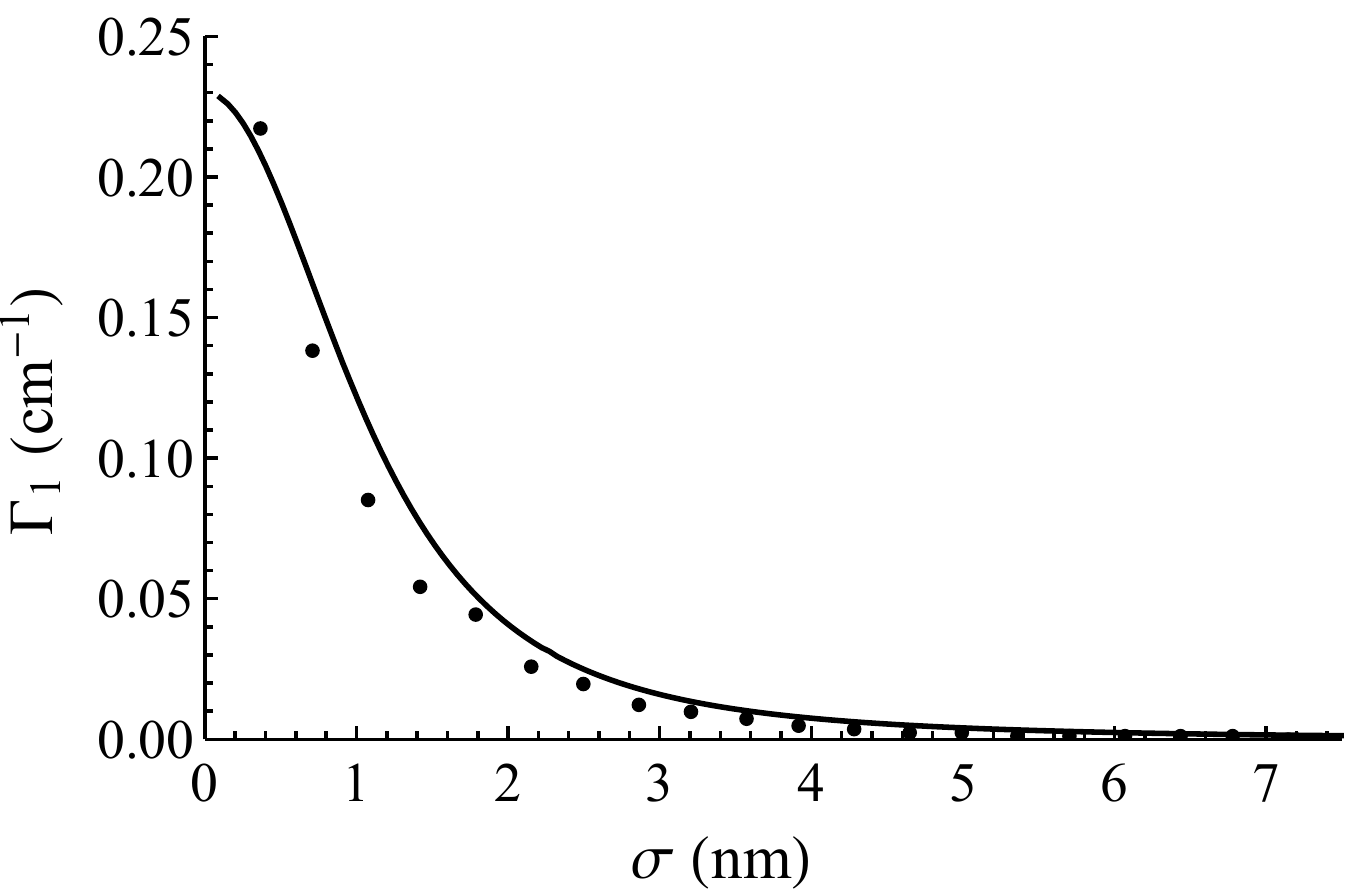}\label{figure}
  \caption{\label{fig_SEKFG} The damping of the first optical phonon eigenmode  $\Gamma_1$ as a function of characteristic scale $\sigma$ of the Gaussian  smooth disorder calculated numerically  with the use of EKFG approach (black dots) and theoretically (black curve) for cubic particles with $L=8.9$~nm. Numerical and analytical results are in good agreement. For $\sigma$ comparable with the particle size the broadening is much smaller that for equivalent point-like impurities. For highest phonon modes the diminishing takes place at even smaller $\sigma$.}
\end{figure}

Utilizing the EKFG approach and studying numerically the Gaussian-correlated disorder introduced in Subsection~\ref{EKFG-D} we arrive to the same conclusions
as in paper~I. Namely, we observe that for a given phonon mode $n$ the broadening essentially depends on the product $q_n \sigma$. When $q_n \sigma \ll 1$ one can use the results for weak point-like impurities, the smooth character of disorder provides only small corrections. Notice that for the first phonon mode condition $q_1 \sigma \ll 1$ implies $\sigma \ll L /(2 \pi) $ which for nanoparticles of nanometer size means that even the disorder correlated on distances of order of several lattice parameters leads to significant suppression of the broadening. This statement is illustrated by Fig.~\ref{fig_SEKFG} depicting the fast decay of $\Gamma_1 (\sigma)$ as $\sigma$ increases observed for cubic particles with $ L \approx 8.9$~nm. In the opposite case  $q_n \sigma \gtrsim 1$ one observes a drastic diminishing of damping. It means that in absence of additional broadening mechanisms this type of disorder is not capable to provide level overlaps leaving the spectrum in separated regime.

\subsection{Lineshift and Mesoscopic Smearing}
\label{SStat}

Disorder yields additional contribution to the broadening of the Raman peak which appears even in the ensemble of {\it identical} disordered  particles due to ``mesoscopic'' smearing of phonon lines. The origin of this smearing could be clarified as follows. In the ensemble of identical particles disorder (say, local mass variations) generates a size independent shift of the maximal phonon frequency $\omega_0$ proportional to the mean mass variation $\langle \delta m \rangle $ and to the impurity concentration $c_{imp}$. For disorders with zero mean this shift is equal to zero. Nevertheless, even in the latter case there exist  fluctuations of the mean (over the particle) impurity mass value due to difference of disorder realizations in various particles. This difference generates different fluctuation-induced shifts of $\omega_0$ in particles.
Upon disorder averaging (over ensemble) these shifts lead to the finite linewidth of a phonon mode as well as to its {\it size dependent} shift. The latter shift is nonzero even when $\langle \delta m \rangle = 0$ because it is proportional to the autocorrelator $\langle ( \delta m )^2 \rangle $, the quantity related to  variance of the function rather to its mean value.

If disorder realizations in various particles are independent, the abovementioned fluctuations obey the Poissonian statistics in the discrete ensemble of particles or the Gaussian statistics in the (quasi)continuous one. This smearing mechanism is similar to the smearing that occurs due to the real particle size variation in powders but appears even for identical particles, stemming from fluctuations of the number of impurities  $N_i$ rather than the number of atoms $N$ in a particle.

The above shift is due to (supposedly, independent) fluctuations of disorder. The relative amount of atoms participating in fluctuations is $N_i / N$ and the relative probability of a fluctuation is $1/\sqrt{N_i}$. It yields:
\begin{equation}\label{Est}
\Delta \omega_0 \propto \omega_0 \, \frac{N_i}{N} \, \frac{1}{\sqrt{N_i}} \propto \omega_0 \, \frac{\sqrt{c_{imp}}}{ L^{3/2}},
\end{equation}
which resembles the linewidth behavior for separated levels. In the overlapped regime the levels start to cross-talk, and the analysis becomes more tricky. Evidently, the effect disappears in the bulk limit $N \to \infty$.

Now let us provide some details.
In the presence of impurities with concentration $c_{imp} \ll 1$ and masses $m + \delta m$ the average atomic mass in a particle reads
\begin{equation}
    \langle m_l \rangle = m + c_{imp} \, \delta m,
\end{equation}
while its variance is
\begin{equation}
    \langle m^2_l \rangle - \langle m_l \rangle^2 = c_{imp} (\delta m)^2.
\end{equation}
When calculating how the mass is distributed in a particle containing $N$ atoms, the latter quantity gives the standard deviation
\begin{equation}
    \Delta m = \delta m \, \sqrt{\frac{c_{imp}}{N}}.
\end{equation}
Since $\omega^2_0$ is inversely proportional to the reduced mass of a cell we have the standard deviation of $\omega_0$ in the form:
\begin{equation}
    \Delta \omega_0 = \frac{\omega_0}{4} \, \frac{\delta m}{\langle \, m_l \, \rangle} \, \sqrt{\frac{c_{imp}}{N}}.
\end{equation}
This quantity is simply related to the disorder strength parameter $S$ introduced above:
\begin{equation}\label{Shift}
    \Delta \omega_0 =  \frac{\omega_0}{4} \, \sqrt{\frac{S}{N}}.
\end{equation}
Again, we see that the frequency shift given by Eq.~\eqref{Shift} is not only proportional to $1/L^{3/2}$ but also depends on other parameters on the same fashion as the linewidth for separated levels. Hence, it should be taken into account. Estimates reveal that it provides about $1/6$ of the overall broadening. For strong disorder and/or overlapped levels this contribution is found to be much smaller.

\begin{figure}[t]
\centering
  \includegraphics[width=0.80\linewidth]{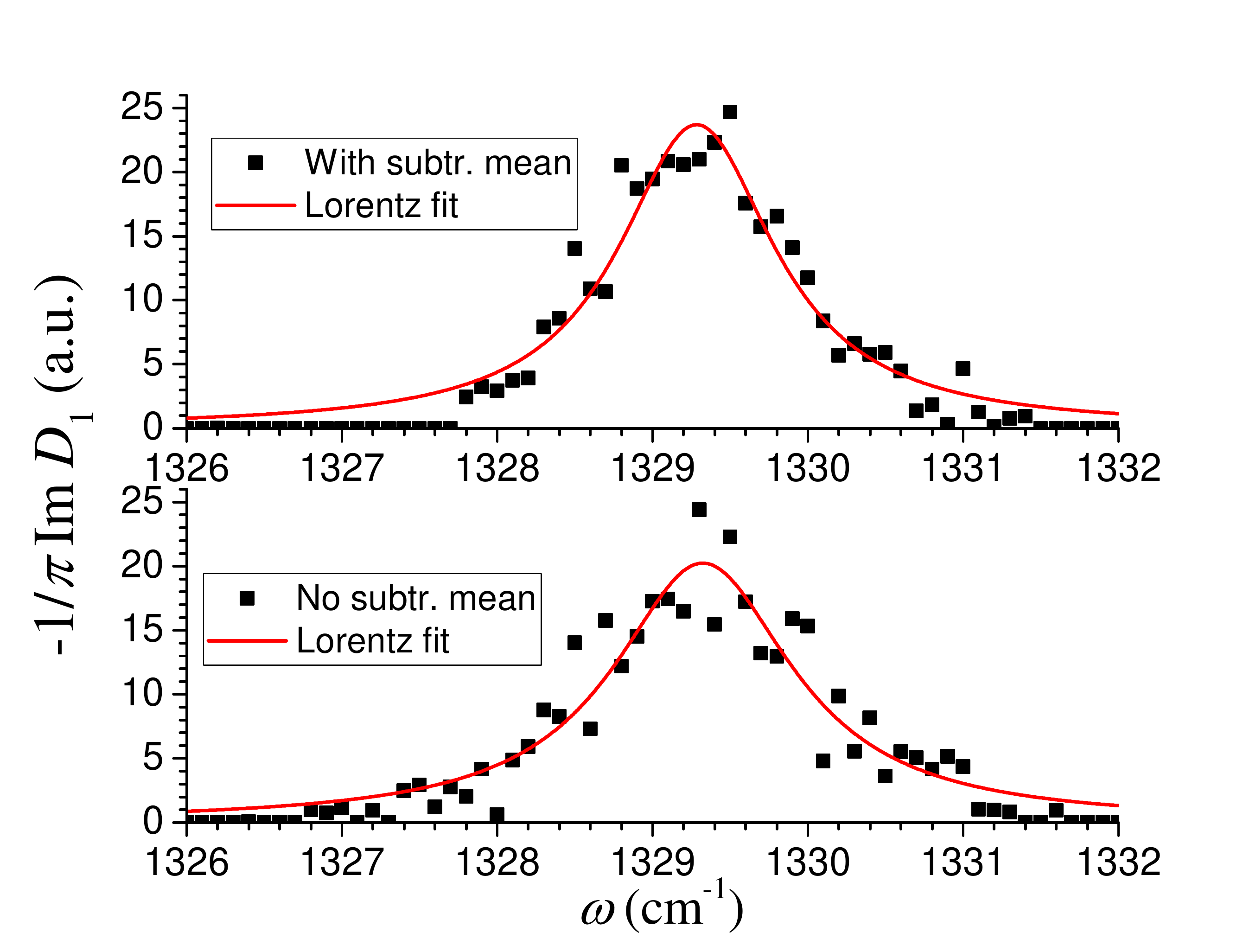}\label{figure}
  \caption{\label{fig_subtr} We investigate the effect of ``mesoscopic'' smearing examining the linewidth of the first phonon mode $\Gamma_1$ for separated levels in 3~nm spherical nanodiamonds with the use of DMM method. In the upper panel the Gaussian disorder is forced ``by hands'' to have zero mean value in each particle separately, which delivers the linewidth $\Gamma_1 \approx 1.2 \, \mathrm{cm}^{-1}$. In the lower panel we did not use that trick allowing the impurity mass  fluctuate from particle to particle. The resulting broadening increases to the value  $\Gamma_1 \approx 1.4 \, \mathrm{cm}^{-1}$.}
\end{figure}

The above arguments are supported by our numerics. In Fig.~\ref{fig_subtr} we show the difference between the broadening of the first phonon mode $\Gamma_1$ in 3~nm spherical diamond particles subject to weak point-like disorder with and without subtraction of the mean impurity mass value for every particle [see Eq.~\eqref{dissubtr}]. We obtain $\Gamma_1 \approx 1.2 \, \mathrm{cm}^{-1}$ for the former case and $\Gamma_1 \approx 1.4 \, \mathrm{cm}^{-1}$ for the latter one.

To conclude Section IV, we check numerically analytical predictions of paper I about weak disorder.
For point-like impurities the linewidth dependencies $\Gamma_n \propto \sqrt{S}/L^{3/2}$ and  $\Gamma_n \propto S/L$ are confirmed numerically and attributed to  regimes of separated and overlapped levels. We corrected the numerical prefactor in $\Gamma_n$ for separated levels, the discrepancy stems from difference between semicircle and Lorentzian descriptions of the lineshape. The phonon linewidth for overlapped levels is found to be growing with the phonon quantum number but slower than the theory predicts. We estimate the crossover scale between separated and overlapped regimes for realistic values of parameters and found that it belongs to nanometer range. For separated levels we reproduce numerically significant diminishing of damping for smooth disorder in comparison with the point-like one even for disorder scales $\sigma$ of the order of several lattice parameters. We also investigate analytically and numerically the ``mesoscopic'' smearing of distribution function which occurs even in ensembles of identical disordered particles, an issue not addressed in paper I.


\section{Results: Strong impurities}\label{StrongR}

In this Section we present the results of numerical modelling for strong disorder and compare them with analytical predictions of paper I. In Subsection~\ref{StrongPrelim} we explain the physical reasons to distinguish between ``weak'' and ``strong'' impurities and outline the results of paper I for strong impurities.  Subsection~\ref{StrongNV} is devoted to numerical study of crossover from weak to strong regime and to the phenomenon of resonant scattering. In Subsection~\ref{ResonLocal} we observe and investigate  strong dependence of the capability for impurity to localize vibrational modes on its location inside the particle, the problem not addressed in paper I because of its analytical complexity. Throughout this Section we assume that the disorder is strong if the condition $S \lesssim 1$ is fulfilled but at least one of the requirements, $c_{imp} \ll 1$ or $| \delta m| / m \ll 1$, is relaxed.

\subsection{Preliminary Remarks}\label{StrongPrelim}

Weak (and dilute) disorder studied in previous Section distinguish from others two important features. First, it is sufficient to use as its measure a single small parameter $S \ll 1 $ (``disorder strength''). Second, parameter $S$ is a product of dimensionless impurity concentration $c_{imp}$ and dimensionless randomness parameter (atomic mass, in our case) squared, $(\delta m / m)^2$, both quantities assumed to be {\it independently} small, $c_{imp} \ll 1$ and $|\delta m | / m \ll 1$.

 When {\it any} of these parameters becomes of order of unity, the physical picture changes, even though the smallness of another parameter provides  $S \ll 1$. For instance, when $c_{imp}$ becomes of order of unity (more carefully, when the phonon mean free path $l_{ph} \sim \sqrt[3]{c_{imp}}$ becomes of order of a few interatomic distances), the approximation of phonon scattering by isolated impurities breaks down, and the {\it multi-impurity} processes which include the interference of phonon scatterings off several impurities come into play. In the lack of detailed theory we touch this issue slightly detecting the crossover to a novel regime at $l_{ph} \sim 3 a_0$ and speculating on general properties of $\Gamma_n (c_{imp})$ for unitary impurities.

On the other hand, when the variation of the random parameter becomes of order of its mean value, $|\delta m|/ m \sim 1$, or even more, the processes of {\it multiple scattering off the same impurity} become important, and  parameter $S$ defined above looses its meaning, the results begin to depend on $c_{imp}$ and $\delta m / m$ (more precisely, on $U$) separately.
Moreover, the physics starts to vary with the sign of $U$: at positive $U$ (heavy impurities) there is no chance to form the long-living optical phonon-impurity bound state. On the contrary, at negative $U$ (light impurities) limited by the condition $\delta m/m \geq -1$ (here equality stands for  vacancy) the impurity scattering is enhanced and even acquires resonant character at certain $U_{min}$. For a vacancy we get $U \to - \infty$, and the mass drops out from the result.

\begin{figure}[t]
\centering
  \includegraphics[width=0.9\linewidth]{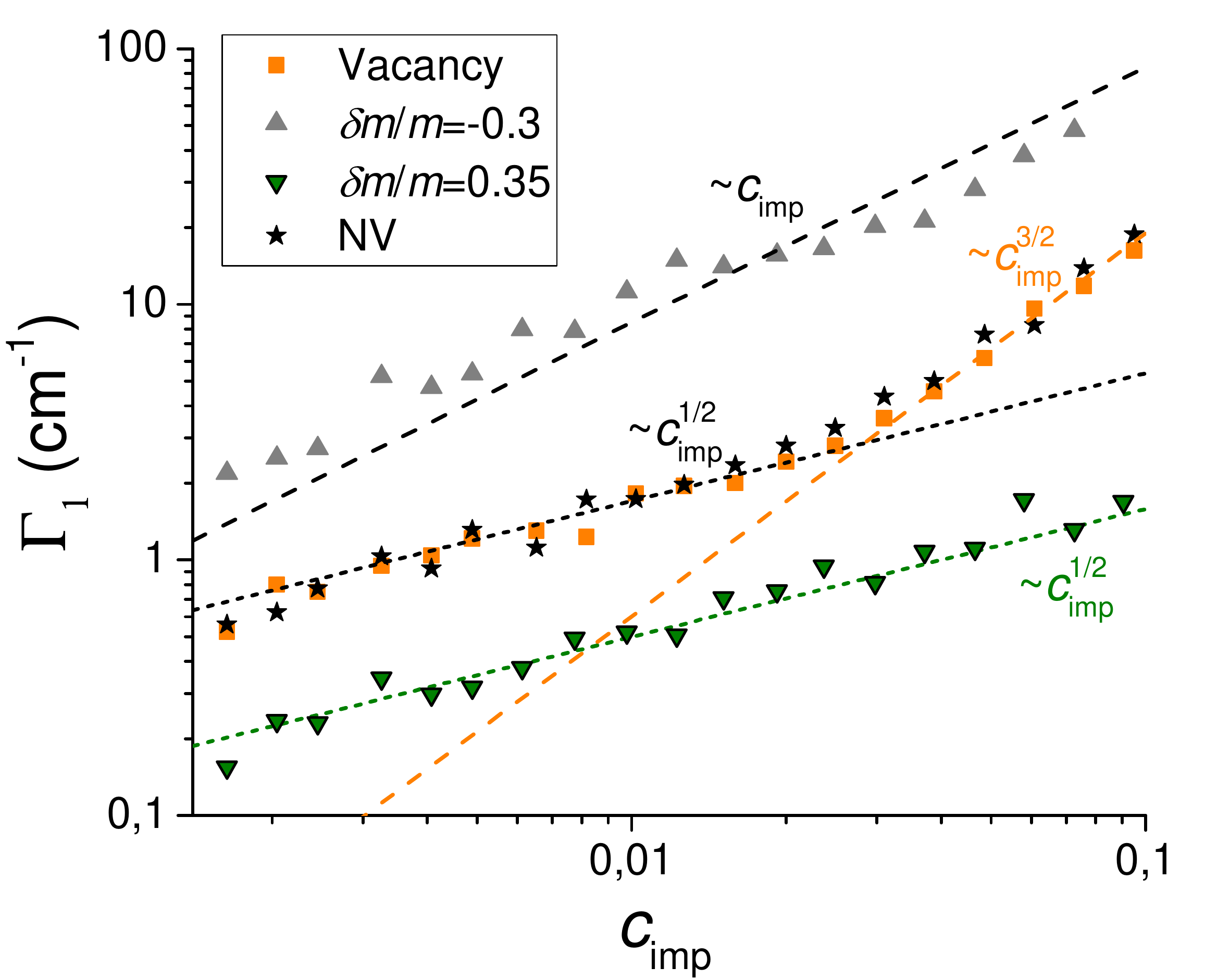}\label{figure}
  \caption{\label{fig_Tc} The  linewidth of the first phonon eigenmode $\Gamma_1$  versus  impurity concentration $c_{imp}$ for various values of the impurity potential $U$, the results  are obtained numerically for 3~nm spherical diamond particles with the use of DMM approach. For heavy impurities (inverted green triangles) the
  broadening follows $\Gamma_1 \propto \sqrt{c_{imp}}$ dependence, whereas light resonant impurities  (grey triangles) reveal  $\Gamma_1 \propto c_{imp} $ dependence. Vacancies (orange squares) demonstrate a crossover from square-root dependence  to the new regime $\Gamma_1 \propto c^{3/2}_{imp}$ at higher concentrations. NV centers (black stars) behave similar to  vacancies.}
\end{figure}

Sketching here the analytical results of paper I for strong disorder, let us mention that the phonon damping $\Gamma_n$ as a function of particle size $L$ and concentration $c_{imp}$ basically follow the same $\sqrt{c_{imp}}/L^{3/2}$ and $c_{imp}/L$ dependencies in the regimes of separated and overlapped phonon levels which it demonstrates for weak disorder. The only difference occurs in the  proximity of resonance wherein appearance of additional long spatial scale $\zeta$ leads to the crossover in $L$-dependence taking place at $L \sim \zeta$. Furthermore, the capability of strong light impurity to capture the phonon  and the frequency of the impurity-phonon bound state evaluated analytically depend on parameter $\zeta$ but not on the location of impurity.

\subsection{Strong Impurities. NV Centers}\label{StrongNV}


\begin{figure}[h]
\centering
\includegraphics[width=0.96\linewidth]{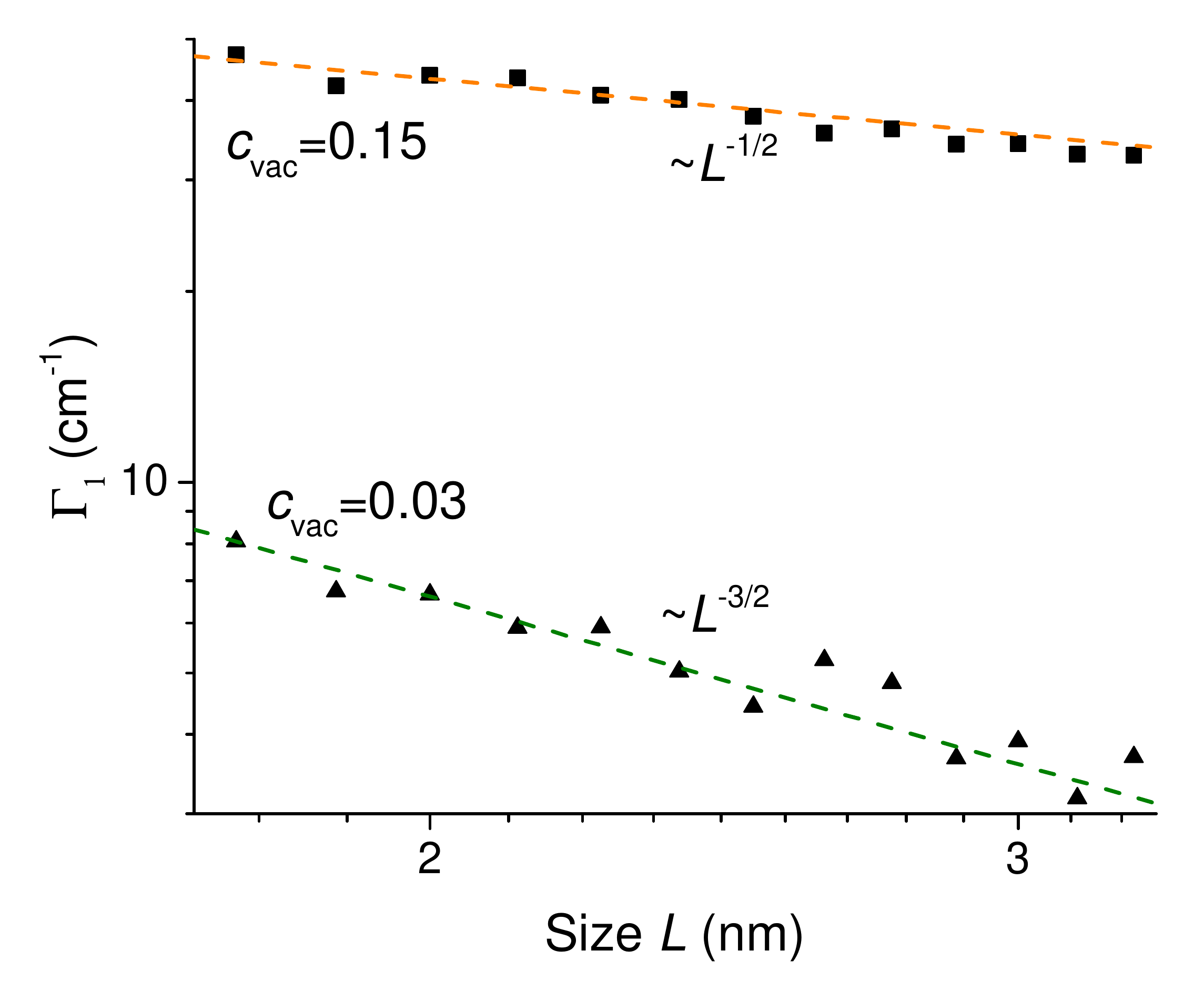}\label{figure}
\caption{\label{fig_Tl} The linewidth of the first phonon eigenmode $\Gamma_1$ as a function of particle size $L$ calculated numerically  with the use of DMM method for spherical diamond particles and impurities in the form of empty vacancies. The plot is presented for two values of vacancy concentration corresponding to ``dilute'' and ``dense'' regimes wherein the linewidth behaves as $\Gamma_1 \propto \sqrt{c_{imp}} / L^{3/2} $ and $ \Gamma_1 \propto c^{\, 3/2}_{imp} / \sqrt{L} $, respectively.}
\end{figure}

Numerically calculated phonon linewidth of the first vibrational mode $\Gamma_1$ as a function of impurity concentration $c_{imp}$ is depicted in Fig.~\ref{fig_Tc} for several values of parameter $U$ (or  $\delta m /m$), namely, (i) for moderately heavy impurities with $U \approx 0.23$
 ($\delta m / m \approx 0.35$) by inverted green  triangles;
(ii) for resonant impurities with $ U \approx -0.43 $ ($\delta m / m \approx -0.30 $)
by grey triangles; (iii) for empty vacancies, which corresponds to the unitary limit $U \to -\infty$ ($\delta m / m \to -1$) by orange squares, and (iv) for NV centers described as a vacancy plus the neighboring heavy impurity with $U \approx 0.14$   ($\delta m/m = 0.17$) by black stars. We observe that in the considered range of concentrations heavy and light (resonant) impurities lead to dependencies peculiar for separated ($\Gamma \propto \sqrt{c_{imp}}$) and overlapped ($\Gamma \propto c_{imp}$) regimes, respectively, which is not a surprise because the theory predicts for resonant impurities the strong enhancement of  prefactor (also seen in Fig.~\ref{fig_Tc}) capable to transfer the system from one regime to another even though the concentration $c_{imp}$ and the absolute value of the mass defect $|\delta m /m|$ are not very different in these cases.

Another interesting phenomenon we observe for unitary impurities and NV centers. As we know from the theory (see also below) dependence on parameter $U$ disappears in the unitarity. Moreover, in Fig.~\ref{fig_Tc} the crossover from the square-root $c_{imp}$-dependence of separated levels to yet another regime $\Gamma_1 \propto c_{imp}^{3/2}$ is seen at high concentrations. This regime is not predicted by our analytics in paper I. Notice that the crossover takes place at $c_{imp} \simeq 0.03$ which presumably coincides with the boundary between ``dilute'' and ``dense'' regimes for the phonon mean free path $l_{ph}$, where the multi-impurity physics begins to be important.

In order to cross-check our understanding of the crossover for $\Gamma_1 (c_{imp})$ depicted in Fig.~\ref{fig_Tc}  we calculate numerically $\Gamma_1 (L)$ dependence for vacancies (see Fig.~\ref{fig_Tl}). We observe that in the
regime when $\Gamma_1   \propto \sqrt{c_{imp}}$ its size dependence is $\Gamma_1 \propto 1/L^{3/2}$, in agreement with our analytics for separated levels. In the novel ``dense'' regime $\Gamma_1 \propto c^{\, 3/2}_{imp}$ it behaves as $\Gamma_1 \propto 1/\sqrt{L}$ yielding
\begin{equation}\label{Dence}
\Gamma_1 \propto \frac{c^{3/2}_{imp}}{\sqrt{L}}.
\end{equation}
Inspecting all formulas for $\Gamma_n$  we discover that each power of concentration always comes with the first power of the mean particle size, i.e., as a product $c_{imp} \times L$.

\begin{figure}[t]
\centering
\includegraphics[width=0.7\linewidth]{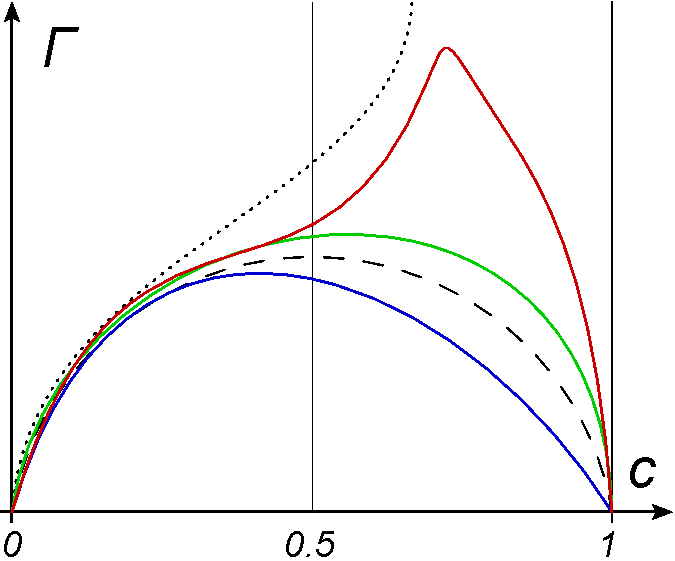}\label{figure}
\caption{\label{Specul} The phonon linewidth $\Gamma$ plotted versus impurity concentration $c$, schematic plot in the entire interval $c \in [0,1]$ for binary disorder, all curves are normalized to the mass of ``first'' material. Blue and green curves depict situations, when the ``impurities'' are just slightly heavier and lighter than the host material, respectively, and the black dashed line represents their separatrix  $\sqrt{c \, (1-c)}$. Black dotted line portrays the percolation transition that occurs in the unitary limit $U = - \infty $ , and the red one shows $\Gamma$ close to criticality, i.e.,  for very light ``impurity'' atoms.}
\end{figure}

At the moment, we have no detailed theory for the behavior given by Eq.~\eqref{Dence} although the underlying physics is evident. Below we just speculate about
the {\it binary} disorder which is invariant under the duality transformation $c_{imp} \, \Longleftrightarrow  \, 1- c_{imp}$.
More accurately, duality condition requires for physical observables the symmetry property (cf. Ref.~\cite{hass1992lattice})
\begin{equation}\label{Dual1}
M_{1}^{A} \, M_{2}^{B} \, {\cal G} \, (c_{imp}) = M_{2}^{A} \, M_{1}^{B} \, {\cal G} (1 - c_{imp}),
\end{equation}
where $M_1$ and $M_2$ are the masses of atoms of the first and second sort, respectively, while $A$ and $B$ are certain exponents (which may coincide or be equal to zero for some quantities) specific for any observable. Transferring all mass dependence to the r.h.s. of Eq.~\eqref{Dual1} we get:
 \begin{eqnarray}\label{Dual2} \nonumber
 {\cal G} (c_{imp})  &=&  {\cal G} (1 - c_{imp}) \, \left(1 + \delta m/m \right)^{\alpha} \\
   &\approx&  {\cal G} (1 - c_{imp})  \, \left(1 + \alpha \, \delta m /m \right),
\end{eqnarray}
where $\alpha = A-B$, and the approximate equality in Eq.~\eqref{Dual2} holds for small $\delta m /m $. Thus, we obtain that mass-dependent observables (including the broadening parameter $\Gamma_n$) are not simply a subject of duality condition $c_{imp} \, \Longleftrightarrow  \, 1- c_{imp}$ but should be simultaneously re-weighted with some mass-dependent prefactors.

The behavior of $\Gamma (c_{imp})$ in the entire interval of concentrations $0<c_{imp}<1$ is schematically plotted in Fig.~\ref{Specul}. It is assumed that the region of small $c_{imp} \ll 1$ corresponds to small amount of impurities with mass $M_2$ and  material with mass $M_1$ forming the host lattice there,
while the region where $c_{imp} \simeq 1$ describes the opposite situation. The entire picture is normalized to the first mass so the mass of the second element is treated as
light or heavy relative to the first one. The linewidth dependence on the parameter $c_{imp}$ for second atoms slightly heavier and slightly lighter than the first ones is depicted  in Fig.~\ref{Specul} by featureless blue and green curves, respectively. Both these curves behave as $\sqrt{c_{imp}}$ near zero and as $a \sqrt{1 - c_{imp}}$ near unity, where $a$ is some mass-dependent prefactor [cf. Eq.~\eqref{Dual2}] with $a < 1$ for heavy $M_2$ atoms and $a > 1$ for light $M_2$ atoms. We match these asymptotes at intermediate concentrations following continuity reasons. With increasing of the second mass the blue curve in Fig.~\ref{Specul} does not change drastically, its right shoulder continue to decrease monotonically. It is not the case for light $M_2$ atoms. To understand it better consider the extreme case of vacancies when $M_2$ is equal to zero and the impurity potential $U$ reaches the unitary limit becoming the infinite point-like on-site repulsion. Appearance of such on-site potential means the elimination of this site from the lattice dynamics.  When the amount of eliminated sites is small, they work as conventional strong impurities; however, at certain critical concentration $c_{cr}$ they lock the propagation of vibrational modes, the transition occurs according to percolation scenario (notice that the particle diffusion and the propagation of vibrational modes on fractals belong to the same universality class, see Ref.~\cite{Nakayama1994}). It is natural to assume that phonons become poorly defined (overdamped) excitations before they die, i.e. the phonon rate is the critical quantity in this problem:
\begin{equation}\label{Percol}
\Gamma \propto (c_{imp} - c_{cr})^{- \tau},
\end{equation}
with $\tau > 0$ being some percolation-related critical exponent. The behavior of $\Gamma$ in the unitary limit is shown in Fig.~\ref{Specul} by dashed black line, with the square-root increase at small concentrations crossing over to the critical behavior near the percolation transition. Finally, the red curve in Fig.~\ref{Specul} represents the $c_{imp}$-dependence of $\Gamma$ for light $M_2$ atoms close to the unitarity. It is drawn basing on continuity arguments as an interpolation between the regimes described by green and dotted black curves, and include the region near $c_{cr}$ where the damping of phonons (or, probably, already ``phasons'', see \cite{Nakayama1994}) reveal critical properties although the real localization of vibrational modes does not take place yet.

The important feature common for vacancies and very light $M_2$ atoms is an inflection point on the left shoulder of $\Gamma(c_{imp})$ dependence stemming from necessity to match square-root and critical asymptotes. Although our methods are not well-suited for the treatment of phonon modes at intermediate impurity concentrations (``dense'' regimes) we believe that the departure from $\sqrt{c_{imp}}$ low-concentration dependence of $\Gamma$ to the more fast $c^{3/2}_{imp}$ behavior seen in Fig.~\ref{fig_Tc} for two numerical plots related to light impurities (NV centers and vacancies) revealing this inflection is the strong argument in favor of  picture we presented above basing on general reasons.

\subsection{Resonant Scattering. Localized States}\label{ResonLocal}

\begin{figure}[h]
\centering
  \includegraphics[width=0.8\linewidth]{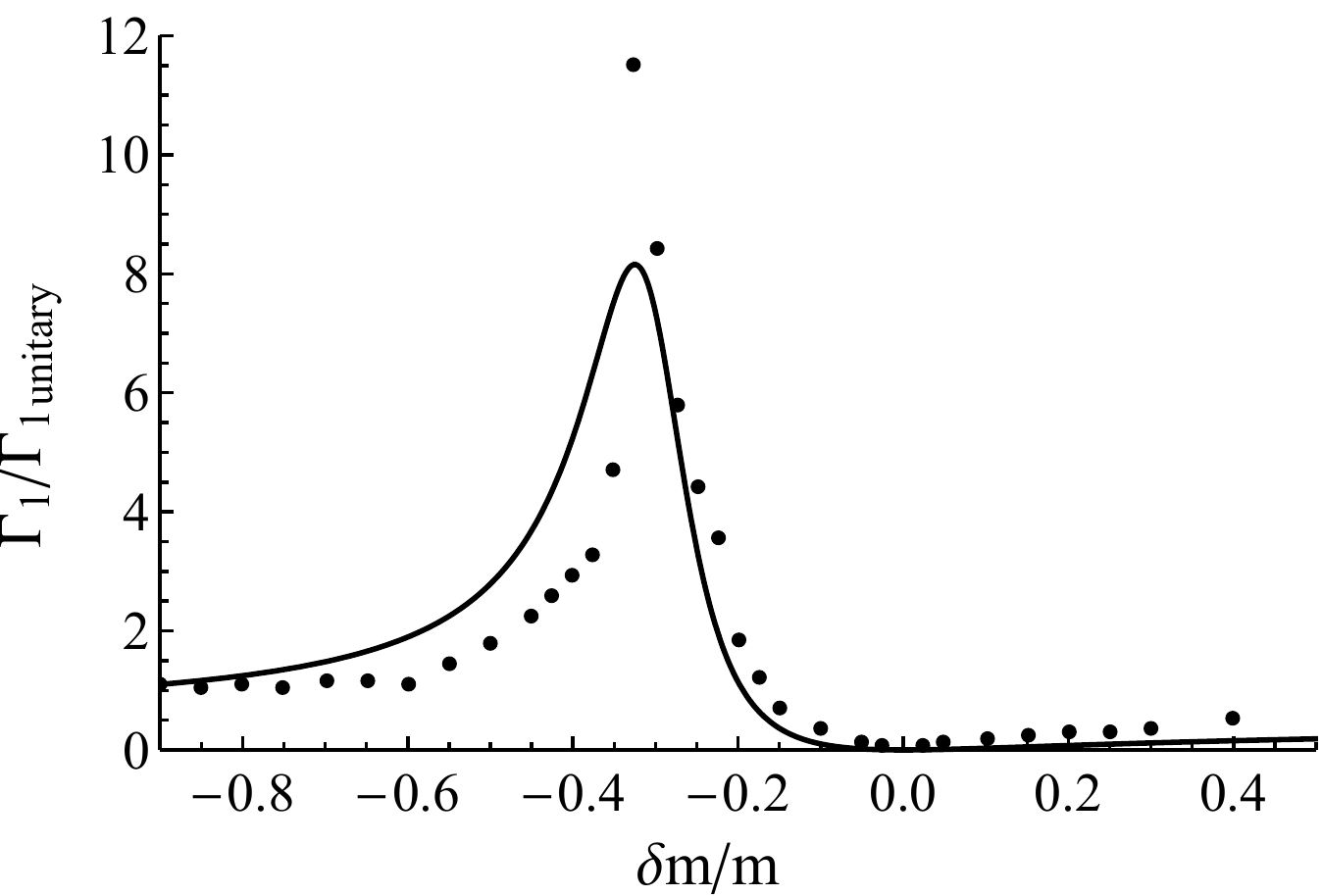}\label{figure}
  \caption{\label{fig_reson} The linewidth of the first optical phonon mode $\Gamma_1$ normalized to its unitary value as a function of $\delta m/m$ at fixed concentration of strong impurities $c_{imp} = 0.02$ calulated for 3~nm diamond particles numerically with the use of DMM method (dots) and analytically using T-matrix approach (line). For $\delta m /m \to -1$ ($U \rightarrow -\infty$) the linewidth saturates (vacancies) and for $\delta m /m \to +\infty$ ($U \to 1$) it vanishes (extremely heavy impurities). At $\delta m /m \approx -0.30 $ ($U \approx -0.43$) the damping acquires resonant character.}
\end{figure}

It was predicted in paper I that the damping of phonons grows essentially as a function of $U$ in the vicinity of its resonant value  determined from the condition
\begin{equation}\label{zeta}
\zeta = \frac{\pi}{2  q_D} \, \left( 1 +  \frac{ 8 \, \pi^2 \, F }{ q_D \, a_0 \, U}   \right)^{-1}
\end{equation}
for relevant spatial scale $\zeta$ to diverge which occurs at
\begin{equation}\label{resonance}
U_{min} = - \frac{ 8 \, \pi^2 \, F }{ q_D \, a_0},
\end{equation}
where $q_D$ is the Debye momentum.  In Fig.~\ref{fig_reson}
we plot  theoretical curve for the linewidth of the first phonon mode $\Gamma_1$ as a function of $\delta m/m$ at fixed $c_{imp} = 0.02$ by blue  line
and compare this plot with $\Gamma_1$ calculated numerically within the scheme presented in Sections \ref{Methods} and \ref{Disorder} of this paper (black dots). We use $q_D$ in as an adjustable parameter to tune the maximum of theoretical curve to coincide with the maximum in our numerics (for reasons to do it see paper I) which
happens at reasonable value $q_D \approx 0.47 \pi/a_0$ yielding $(\delta m / m)_{min} \approx -0.30 $ ($U_{min}\approx - 0.43$).
We see that the analytical theory underestimates the effect. It is expected because the $T$-matrix approximation we used only manifests the phenomenon while for its detailed treatment specific methods adopted to deal with resonant scattering are required. For light impurity atoms with $\delta m / m < -0.7 $ 
$( U < -2)$  and heavy ones with $\delta m /m \to + \infty $ ($U \approx 1$) the damping $\Gamma_1$ obtained numerically saturates at different values in accordance with theoretical predictions. Generally, we report good {\it qualitative} agreement in description of resonant features between the analytics of paper I and the numerical experiment of this paper. Notice that we presented resonant behavior in Fig.~\ref{fig_reson} in terms of $\delta m / m$ instead of $U$  because it looks more evocative.

\begin{figure}[h]
\centering
  \includegraphics[width=0.9\linewidth]{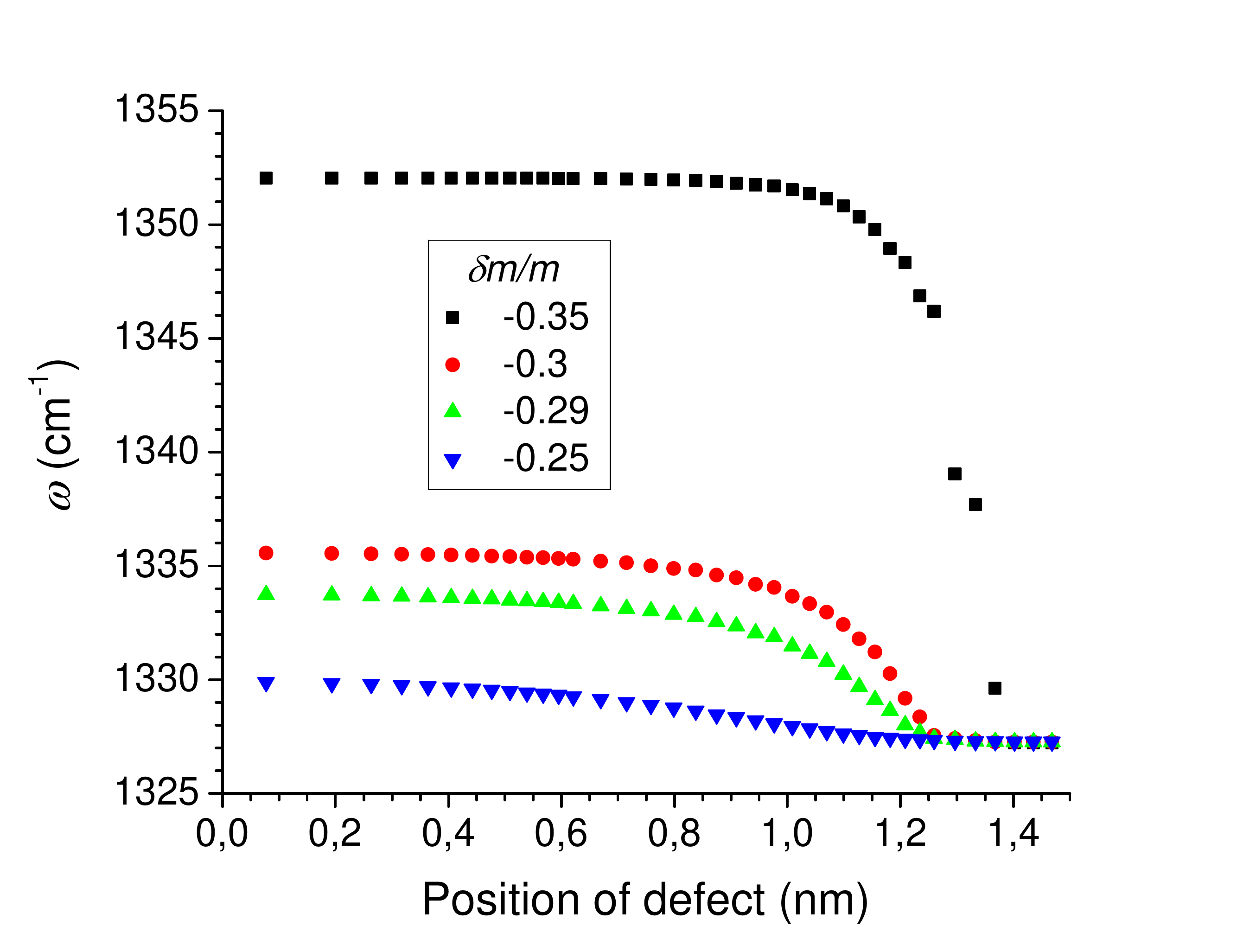}\label{figure}
  \caption{\label{fig_isol} The highest phonon frequency in a particle versus the position of single impurity of a given mass calculated numerically for 3~nm spherical diamond particles with the use of DMM method. When the frequency is higher than $\omega_0 = 1333 \, \mathrm{cm}^{-1}$ the corresponding vibrational mode  is localized on the impurity. The maximal phonon frequency decreases when the distance between the location of impurity and the center of a particle grows. It leads to absence of localization  if the defect is close enough to the particle boundary. }
\end{figure}

\begin{figure}[h]
\centering
  \includegraphics[width=0.8\linewidth]{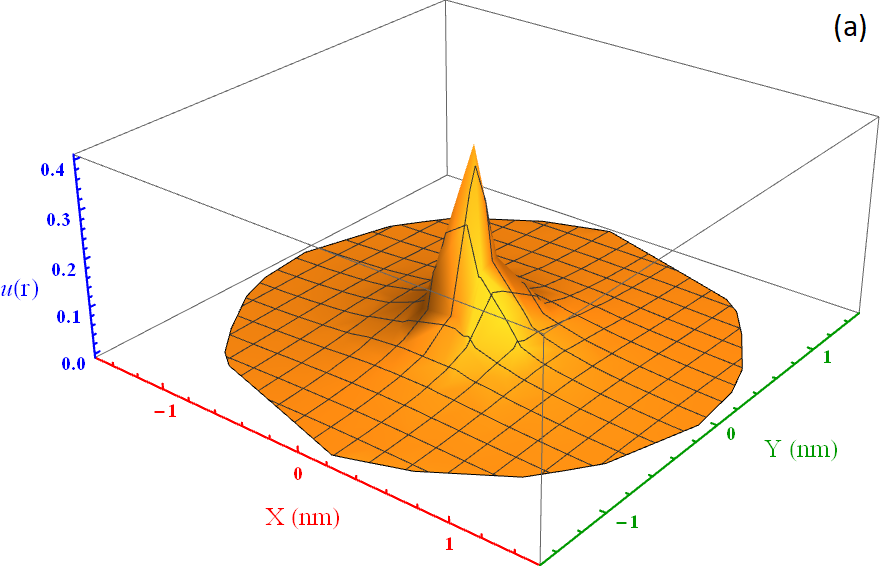}
  \includegraphics[width=0.8\linewidth]{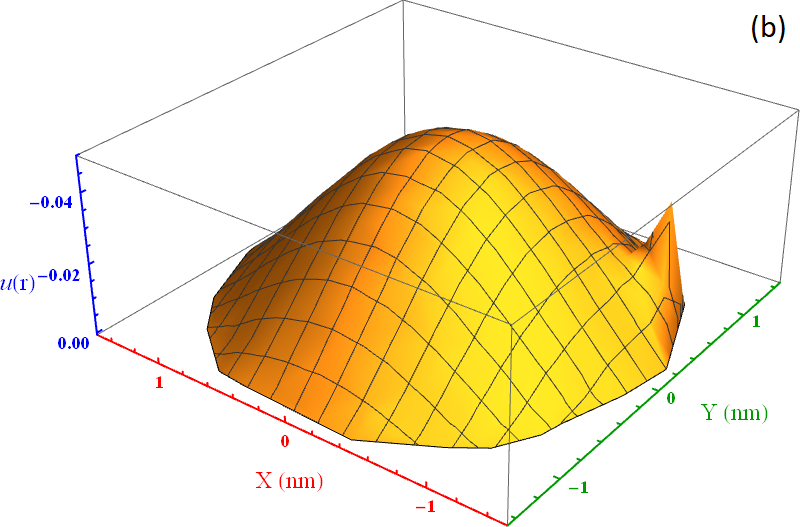}
  \caption{\label{fig_loc} The phonon wave function obtained numerically for 3~nm spherical diamonds using DMM method. Two different locations of impurity with $\delta m /m = -0.33 $ are studied. When the defect lies near the particle center the highest in energy vibrational mode is localized (a), whereas the impurity  settled near the boundary leads to the wave function resembling the pure case with some local feature around the defect (b). }
\end{figure}

 Now we shall analyze a possibility to localize vibrational mode on the impurity. The energy of this state located just above the maximal energy of optical phonon $\omega_0$ is found to behave in paper I as follows:
 \begin{equation}\label{omega_loc}
 \omega_{loc} = \omega_0 \, \left[ 1 + F \, \left(a_0 / \zeta \right)^2 \right].
 \end{equation}

We model the single impurity problem investigating numerically 3nm spherical diamond particles by means of DMM method. We obtain very interesting new phenomenon which has not been predicted by analytical theory of paper I. Namely, we find that the capability for impurity to localize the phonon strongly depends on the location of this impurity inside the particle: it is maximal at the particle center and rapidly decays to the boundary.

Our argument in favor of this picture is presented in Fig.~\ref{fig_isol} where we plot the maximal phonon frequency as a function of distance from defect to the particle center for several nearly-resonant values of the impurity potential $U$. One can see that if the localized state emerges ($U_{min} > U $) its frequency is almost constant as long as its wave function does not ``feel'' the boundary  (see Fig.~\ref{fig_loc}). When the defect is near the boundary $\omega_{loc}$ decreases and the localized state disappears. It occurs because the amplitudes of optical vibrations near the boundary are much smaller than at the center, and the impurity cannot ``catch'' the phonon which results in usual scattering rather than in localization of the vibration.

In Fig.~\ref{fig_loc} we portray the wave function for the highest phonon mode in the localized regime $U_{min} > U$ for two particular cases, namely when the same impurity is located at the center of the particle (first panel) and when it lies near its boundary (second panel), thus visualizing the above reasoning. We see that in the former case the wave function is concentrated in the closest vicinity of the defect, the decay rate being much shorter than the particle size which reflects the phonon localization. On the contrary, if we settle down impurity near the boundary, the phonon wave function is smeared over the particle
which implies an extended state. Moreover, the wave function in the latter case resembles the pure case (cf. Fig.~3 of Ref.~\cite{utesov2018raman}), the only impurity-induced disturbance is seen in the neighborhood of the impurity.

To summarize the results of this Section, we investigate numerically the phonon linewidth for nanoparticles subject to strong dilute disorder with $|U| \gtrsim 1 $ and $c_{imp} \ll 1$. The regimes of separated and overlapped levels observed for weak dilute disorder are found to survive in this case, as well. Inspecting the phonon broadening for vacancies and NV centers at yet higher concentrations we identify the crossover to the regime with $\Gamma \propto c^{3/2}_{imp} /\sqrt{L}$ behavior not predicted by the theory of paper I, and attributed this regime to multi-impurity scattering processes and proximity to percolation transition. We sketch qualitatively $\Gamma (c_{imp})$ dependence for binary disorder at arbitrary concentrations $0 < c_{imp} < 1$ using duality arguments and suggesting the critical behavior for the linewidth parameter in percolation scenario. Our studies of resonant regime of damping reveal that the numerics confirms all principal predictions  of the theory, although the latter underestimates the amplitude of the effect.
At last, we confirm numerically the formation of phonon-impurity bound states described in paper I, and observe that the capability of strong light impurity to localize the phonon mode varies with its position inside the particle monotonically decreasing from the particle center to its boundary. The latter fact throws a bridge  to our analysis of surface defects in nanoparticles presented in the next Section.

\begin{figure}[t]
\centering
  \includegraphics[width=0.9 \linewidth]{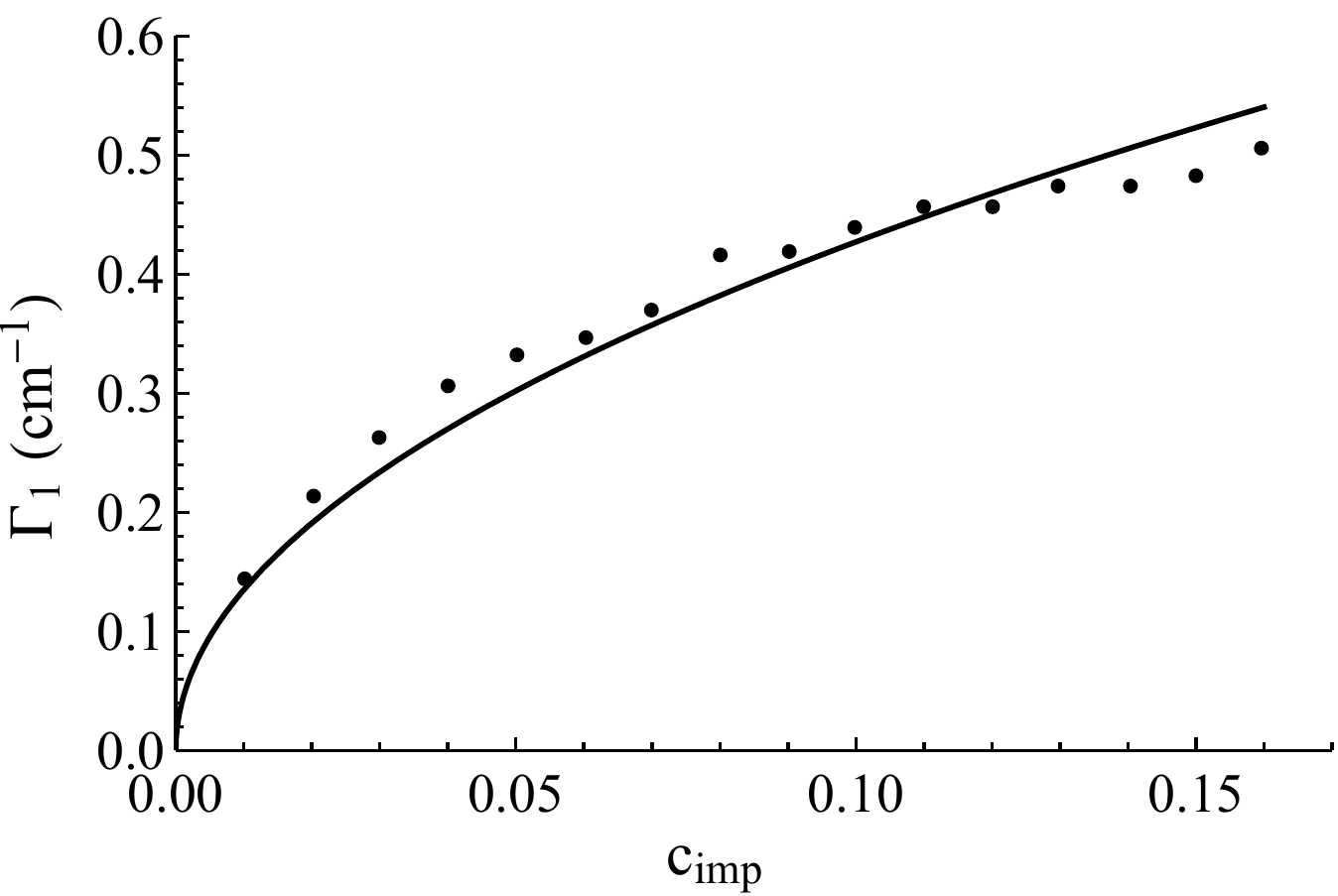}\label{figure}
  \caption{\label{fig_Gbites} The linewidth of the first phonon line $\Gamma_1$ as a function of surface impurity concentration $c_{imp}$ calculated with the use of EKFG approach for  diamond 4.5 nm particles within the model of nibbled apples. Numerics (dots) is well-described by the square-root dependence (line).}
\end{figure}


\section{Results: Surface Corrugations}\label{Surface}

In this Section we present our results concerning the influence of several types of realistic surface disorder on the broadening of volume optical phonon modes which contribute to the Raman spectrum of nanoparticles.

Below we are interested in the effect of $\it surface$ (i.e., two-dimensional or quasi two-dimensional)
irregularities on the behavior of {\it volume} (i.e., three-dimensional) excitations (volume phonons). Surely, both propagating surface modes (surface phonons) and surface-volume mixed modes (breathers, etc.) exist and play important role in the physics of nanoparticles. Moreover, approaches we developed in present work are applicable for treatment of these modes, either. However, the characteristic frequencies of surface modes lie far away from the frequency range relevant for the main Raman peak. Therefore, we postpone their investigation for the future.

\begin{figure}[h]
\centering
  \includegraphics[width=0.9 \linewidth]{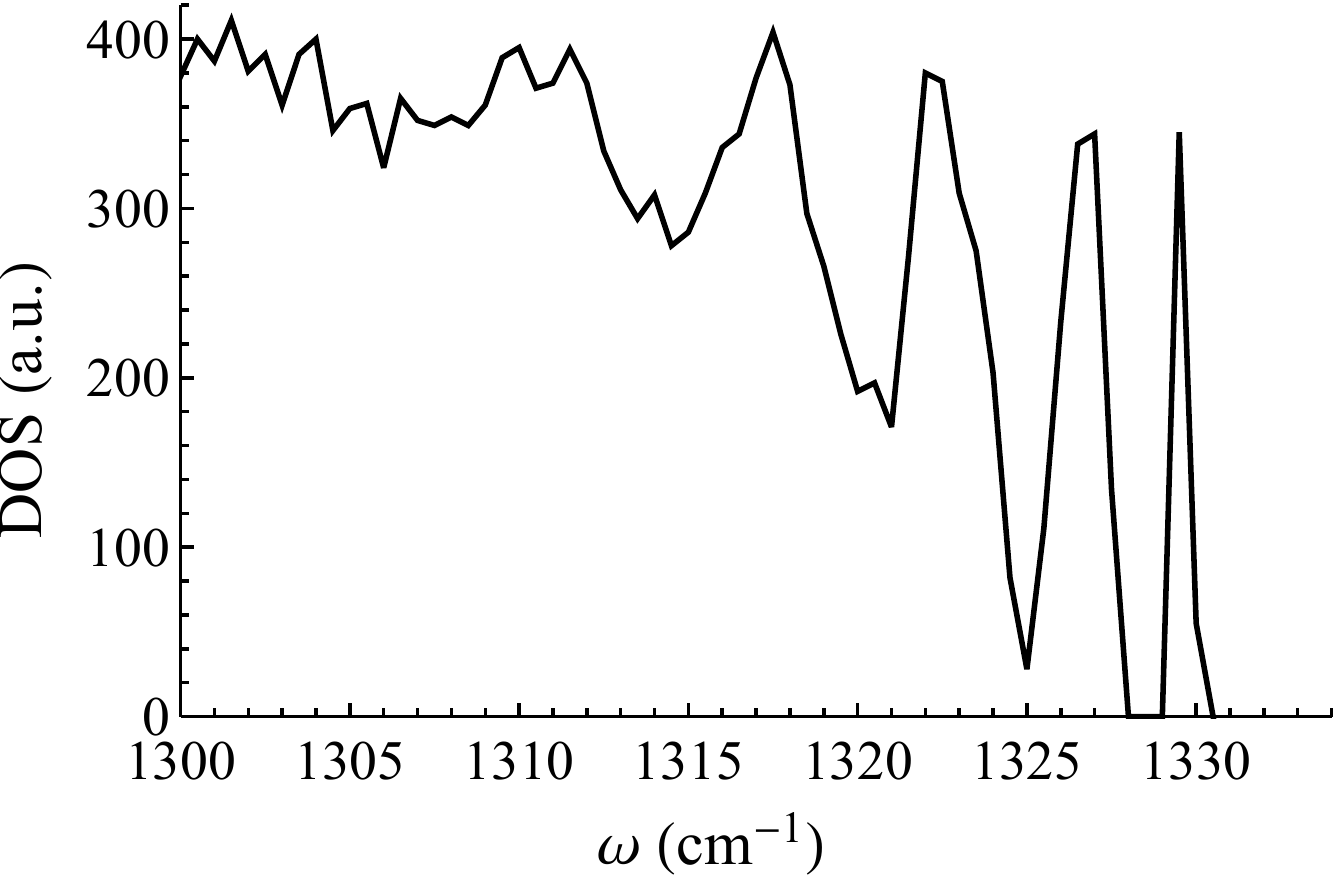}\label{figure}
  \caption{\label{fig_SDos} DOS  calculated  numerically for disordered 4.5~nm diamond  particles with the use of EKFG approach within the peeled apples model (result for nibbled apples is very similar). A bunch of lowest phonon levels are separated whereas higher levels  overlap due to the same amount of surface disorder.}
\end{figure}

When investigating surface corrugations, we observe only minor influence of this type of disorder on the broadening of the first phonon mode as compared to the volume disorder considered in previous Sections, the statement valid for both ``peeled apples'' and ``nibbled apples'' models we studied.
Even for strongest disorder we get $\Gamma_1 < 1 \, \text{cm}^{-1}$.  In Fig.~\ref{fig_Gbites} we plot the broadening parameter $\Gamma_1$ versus impurity concentration $c_{imp}$ within the framework of the nibbled apples model. The volume of a particle is taken clean, and the surface disorder fails to overlap the first level with its neighbor. It leads to the square-root dependence of the linewidth on the impurity concentration seen in Fig.~\ref{fig_Gbites}, which agrees with the prediction of paper I for separated levels. Notice that the notion of concentration should be revised for surface impurities. Say, for nibbled apples one should count the number of ``jobbsings'' (removed bits) and then divide it by the total number of surface bricks whereas for peeled apples which could be regarded as convex irregular polyhedra the definition of concentration includes the average distance between the ribs in disordered particle divided by the same quantity in the pure one.


The importance of surface disorder increases on higher phonon levels where the notion of classical chaos is better applicable since $1/n$ is a version of  quasiclassical parameter. For highest modes the (almost) classical chaotization due to surface roughnesses works better resulting in the overlapped regime at the same amount of impurities which did not allow lowest levels to overlap (see Fig.~\ref{fig_SDos}).
We conclude that surface corrugations may slightly affect the main Raman peak only on its left shoulder, the scattering by volume impurities remains the dominant broadening mechanism in nanoparticles.

Another feature of surface disorder which makes it less important than the volume one is the rapid decay of its contribution to $\Gamma_n$ with increasing $L$. We shall demonstrate it considering the nibbled apples model as an example. This model allows two modifications: for
one of them the size of jobbsings is scaled with the particle size and for another one it does not, so for very large particles impurities acquire truly surface point-like character.

In the former case the dependence $\Gamma_n (L)$ could be obtained from scaling arguments. Inspecting the EKFG equations we observe that due to (statistically understood) scale invariance of the disordered EKFG problem with this type of imperfections its eigennumbers scales in the same manner as for pure EKFG problem~\cite{utesov2018raman}:
\begin{equation}
    q^2_n(L_1) \, L^2_1 = q^2_n(L_2) \, L^2_2.
\end{equation}
Thus, ensembles of particles of different sizes are similar to each other, and corresponding broadenings are mutually related by the scaling law
\begin{equation}
    \Gamma_n (L_1) \, L^2_1 = \Gamma_n (L_2) \, L^2_2 \quad \Longleftrightarrow \quad \Gamma_n (L)\propto \frac{1}{L^2}.
\end{equation}
We argue that this $\Gamma_n \propto 1/L^2$ dependence is the slowest size dependence possible in both models of surface corrugations (at least, for separated phonon modes).
Qualitatively, it can be understood as follows. Let us inspect the Born impurity scattering on separated levels [see Eq.~(29) of paper I]. The squared linewidth parameter $\Gamma_n$ is proportional to the fourth power of the phonon wave function $Y_{n} \propto 1/\sqrt{N}$. Let us introduce $L_d$ as characteristic scale of surface disorder. Then each of four wave functions provides  the factor $(L_d/L)/L^{3/2}$ and the surface area where the interaction with disorder occurs is estimated as $c_{imp} L^2$. It yields $\Gamma_n^2 \propto L^4_d/L^8$. If now $L_d$ scales as the particle size, $L_d \propto L$ (first modification of nibbled apples model), than we get $\Gamma_n \propto 1/L^2$. If however the surface disorder is not scaled with the particle size, $L_d \propto const$ (second modification), we obtain $\Gamma_n \propto 1/L^4$. Intermediate situation $L_d \propto L^{\beta}$ with $0 < \beta < 1$ yields $\Gamma_n \propto 1/L^{4 - 2 \beta}$. The general reason for these rapid decay laws is that the phonon wave function in disordered region near the surface tends to zero (cf. with position-dependent localization in Subsection~\ref{ResonLocal}). For disorders scaled slower than $L$ the relative volume of disordered region decreases when the particle growths. It results in faster decay of scattering amplitude.

\begin{figure}[t]
\centering
  \includegraphics[width=0.96 \linewidth]{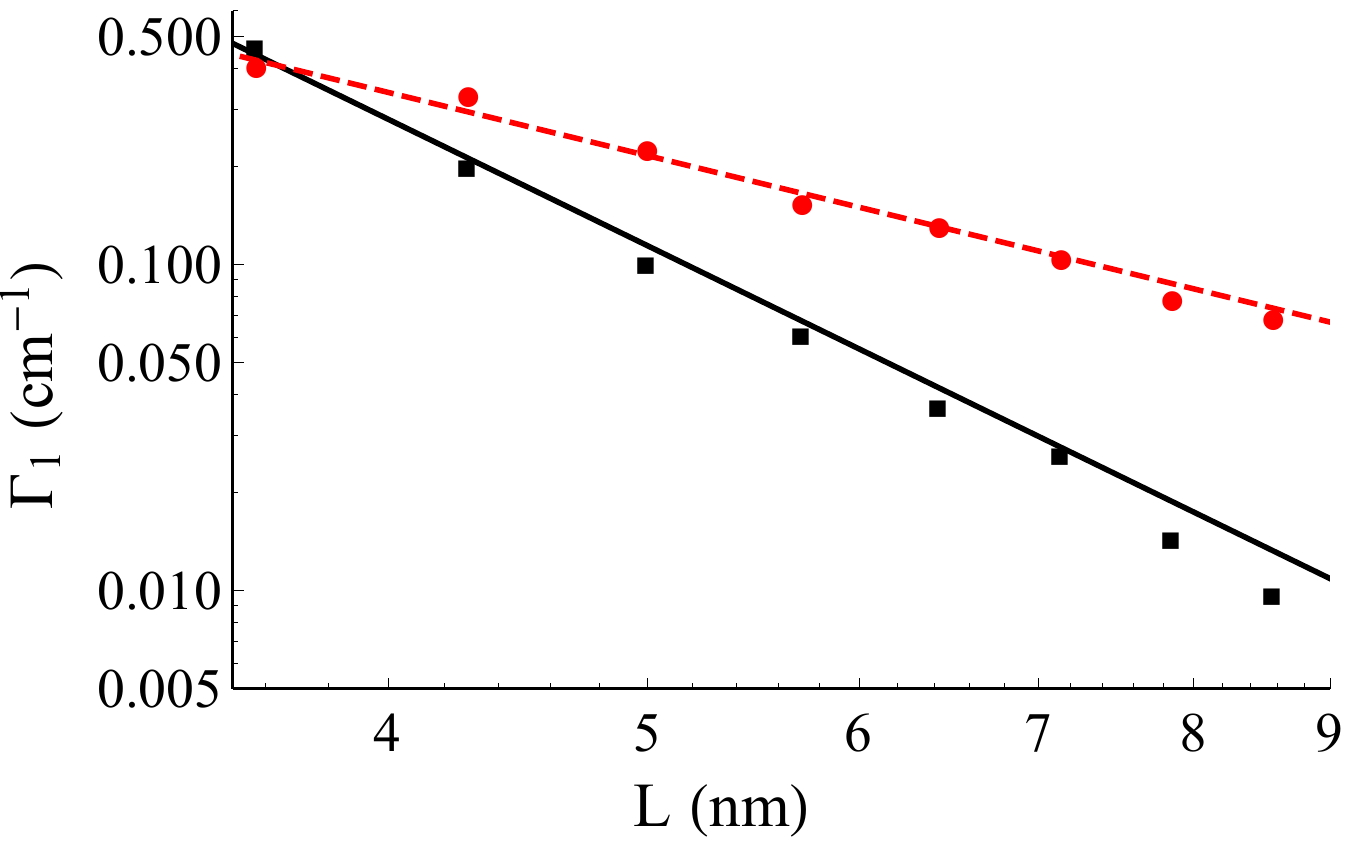}
  \caption{\label{fig_SL} The linewidth of the first optical phonon mode $\Gamma_1$ versus the particle size $L$ calculated numerically for the model of nibbled apples with the use of EKFG method at surface impurity concentration $c_{imp}=0.1$. Red dots correspond to scaled disorder and black squares to the unscaled one, and the lines depict $1/L^2$ and $ 1/L^4$ dependencies, respectively.}
\end{figure}

This is exactly what we observe in our numerics. Result is depicted in Fig.~\ref{fig_SL}, where $1/L^2$ and $1/L^4$ regimes are found for $\Gamma_1$ calculated for the nibbled apples model in its scaled and unscaled modifications, respectively.

In this Section we studied numerically the influence of surface imperfections on lifetimes of volume vibrational modes. We observed that surface corrugations yield minor input to the broadening of phonon levels for volume optical phonons as compared to the volume disorder considered in Sections~\ref{WeakR} and \ref{StrongR}. Focusing on the regime of separated levels we found the concentration dependence of the first phonon mode linewidth in the form peculiar for volume impurities and observe its rapid increase with increasing of the quantum number. The phonon linewidth for surface impurities is shown to decay with $L$ faster that for volume impurities, namely as $1/L^2$ for surface imperfections scaling with particle size and as $1/L^4$ in the the opposite case.


\section{Discussion and Conclusions}\label{DiscConcl}

In this concluding Section we summarize the outcome of numerical experiments reported in paper II and compare them with the results of analytical consideration of paper I in Subsection~\ref{Summary}. Possible applications and generalizations of our theory  as well as  its disadvantages and limitations are discussed in Subsection~\ref{Prospect}.

\subsection{Summary}\label{Summary}

Both papers I and II which constitute the present work are devoted to the treatment of disorder influence on propagating optical vibrational modes in powders of nanoparticles of nonpolar crystals considering diamond particles as representative example. Linewidths and lineshifts of phonon modes as well as the shapes of individual spectral lines in phonon spectra are investigated as functions of mean particle size, particle shape, impurity concentration, strength and type of disorder and the phonon quantum number, addressing these issues analytically in paper I and numerically in present paper. Furthermore, we apply our knowledge of this subject for the analysis of structure of the main Raman peak, in view of the possibility to extract confidently the values of abovementioned parameters from  experimental data~\cite{ourShort}.

In both papers we apply the methods of treatment we developed previously for pure nanoparticles in Refs.~\cite{koniakhin2018raman} and \cite{utesov2018raman}, utilizing discrete atomistic approach in the form of DMM-BPM method and continuous elasticity theory-like approach solving EKFG equations with Dirichlet boundary conditions, depending on particular problem at hands. Parameters emerging in our approaches are simply related to parameters of microscopic models of solids, i.e., of the Keating model. We properly adopted both methods for analytical calculations in paper I and for numerical experiment in paper II.

As far as the disorder is concerned, we investigated several types of imperfections most interesting from theoretical and experimental points of view. In paper I
we studied analytically weak Gaussian disorder in its point-like and smooth modifications and strong binary disorder, both considered in the regime of ``dilute'' concentrations. In paper II we had two objectives, namely to check numerically predictions of the theory and to investigate
realistic regimes of disorder hardly analyzable analytically. Executing the second task we studied disorder at intermediate impurity concentrations and
arbitrary disorder strengths as well as in two models of surface corrugations. We also investigated isotopic impurities and NV centers (nitrogen plus vacancy) widespread in diamonds.

The main observation of paper I is existence of two regimes of behavior for phonon linewidths $\Gamma_n$ depending on the particle size and impurity strength investigated. For smallest particles and/or weakest disorder the phonon levels are separated. It results in $\Gamma_n \propto \sqrt{S} / L^{3/2} $ dependence on these parameters,
the prefactor varying with the particle shape and the phonon quantum number could be calculated either analytically or numerically. When the particle size and/or the strength of disorder increases the levels start to overlap, and the linewidth crosses over to another regime $\Gamma_n \propto S/L$ with different quantum number and shape dependent prefactor.

Investigating numerically weakly disordered particles we confirm analytical predictions of paper I. For point-like impurities we demonstrate the regimes of separated and overlapped levels and reproduce the dependencies $\Gamma \propto \sqrt{S}/L^{3/2}$ for the former and  $\Gamma \propto S/L$ for the latter regime, respectively. We slightly correct the numerical prefactor in the former case and attribute this correction as certain disadvantage of Lorentzian approximation. We observe that the rapid growth of the linewidth as a function of quantum number predicted by analytical theory of paper I overestimates the effect which nevertheless exists and is found to be important for the fit of experimental data. We estimate crossover scales between the regimes and obtain that for realistic values of disorder it lies on the scale of nanometers or dozens of nanometers for the mean particle size $L$, i.e., in the most intensively studied and most interesting range of parameters. Considering the case of a smooth random potential we inspect and confirm numerically significant decrease of the linewidth when the characteristic spatial scale $\sigma$ of the potential grows towards the particle size $L$. We also address (both analytically and numerically) an issue which has been only slightly touched in paper I discussing the phenomenon of mesoscopic smearing of distribution function which is predicted to arise in the ensemble of disordered particles due to variations of disorder realizations in various particles and survive even in the ensemble of identical in size and shape particles.

 We examine numerically the phonon line broadening in the regime of strong rare impurities when the parameter $S$ looses its meaning and the disorder is characterized by two independent quantities such as  the dimensionless impurity potential $U$  and the dimensionless impurity concentration $c_{imp}$, allowing the potential $U$ to be strong enough, $|U| \gtrsim 1 $, and mostly keeping the concentration dilute, $c_{imp} \ll 1$. We find
that  regimes of separated and overlapped levels with their characteristic $c_{imp}$- and $L$- dependencies survive for strongly disordered particles, as well, although the crossover between these regimes could be shifted due to large $U$-dependent prefactor.  For vacancies and NV centers we discover a crossover
 which occurs at intermediate concentration to the regime with $c^{3/2}_{imp}/\sqrt{L}$ dependence of the linewidth not appearing in the low-$c_{imp} $ theory of paper I. It stems from multi-impurity scattering processes (``dense'' defects). We also argued that for binary disorder the abovementioned  dependence [more accurately, existence of an inflection point in $\Gamma_n (c_{imp})$] emerges for light enough impurities reflecting the proximity to percolation transition taking place in the unitary limit, and qualitatively restore the behavior of $\Gamma_n (c_{imp})$ in the entire range of $c_{imp}$.

When examining the resonant enhancement of phonon damping, we observe a good qualitative agreement between the analytical theory and the numerical experiment. Although the former underestimates the amplitude of the effect, it qualitatively reproduces all principal features of the behavior of  damping $\Gamma_n$ as a function of its parameters. Similar agreement we report for our numerical study of phonon localization by a strong light impurity. The novel phenomenon not predicted analytically and investigated only by numerical tools in present paper is the strong dependence on the location of impurity inside the particle observed for the capability of impurity to localize the phonon: this capability is maximal when the defect is located in the center of a particle and decays rapidly when it moves towards the boundary.

We inspect how surface corrugations affect linewidths of volume (optical) phonons studying numerically two models of surface disorder  named the peeled apples model and the nibbled apples model. For reasonable values of parameters we found that the surface contribution to damping is essentially smaller than the contribution of volume  disorder. Since the effect is small we concentrated mainly on the regime of separated levels. The linewidth of the main phonon mode $\Gamma_1$ as a function of surface impurity concentration is shown to behave similar to volume impurities and growths rapidly for highest phonon modes due to their better chaotization. The phonon linewidth decays with increasing of  particle size faster than it occurs for volume impurities as a result of rapid decrease of phonon wave functions on  surface. The character of this decay is estimated analytically and calculated numerically as $\Gamma_n \propto 1/L^2$  for disorder scaled with the particle size and as $\Gamma_n \propto 1/L^4$ for unscaled imperfections.

We monitored numerically asymmetry of phonon lines and their non-Lorentzian shape predicted in paper I.

We conclude that the theory of paper I is verified and approved by the numerics of paper II, the minor deviations caused by
approximate character of analytical approaches could be easily corrected. Furthermore, the numerical methods formulated in paper II allowed to study the regimes which are hardly achievable analytically but interesting from experimental point of view.


\subsection{Discussion and Prospectives}\label{Prospect}




The present paper continues our efforts to build up a new microscopic theory of Raman scattering in nanopowders of nonpolar crystals started in Refs.~\cite{koniakhin2018raman,utesov2018raman,our3,ourShort}. We believe that this theory should replace in usage the phonon confinement model previously applied for the analysis of Raman spectra of nanoparticles. Starting from the microscopic quantum description of mechanical vibrations in finite-size systems and incorporating the photon-phonon interaction under the assumption (obviously valid for nonpolar crystals) that the polarization of a solid occurs due to atomic displacements, this theory, being applied for interpretation of experimental data, is able to provide us with the most complete information about parameters of a powder, shape and structure of particles constituting the powder, and collective excitations that govern atomic dynamics in particles.

All this could be done with accuracy limited only by the accuracy of Raman (optical and therefore very precise) experiment. The fit of Raman experiment performed in Ref.~\cite{ourShort} where four parameters of nanopowder have been confidently extracted from data clearly demonstrates the manifold increase of precision when the data are evaluated within the framework of our method as compared to the PCM approach, the latter is shown to work worse and worse for smaller particles.

In addition, the PCM is in fact a purely phenomenological approach which simply replaces the bulk result by a convolution of bulk result with certain inexplicable Gaussian, with further assignment for the Gaussian-caused decay rate of a phonon spectral weight the meaning of a particle size. Meanwhile, it has been shown in Ref.~~\cite{koniakhin2018raman} that the microscopic theory (in its approximate version) allows the formulation in terms of convolution of bulk formulas; however, the function convoluted with the bulk DOS have nothing in common with the Gaussian. That is why we believe that several attempts made recently in order to cure the PCM are doomed at best to a partial success.
On the contrary, significant increase in the accuracy of data interpretation promised by present approach is of paramount importance for the industrial manufacturing of nanoparticles as well as for their scientific and technological employments.

The only imperfection of this theory in comparison with the PCM is its relative complexity. Indeed, since even the eigenfrequency of the main phonon mode was shown to be 20\% varying with the particle shape~\cite{utesov2018raman}, both DMM-BPM and EKFG methods require numerical calculations (not very tedious, though) for any particle shape beyond the minimal set of cube, sphere and cylinder. To make the things easier, we plan a paper revisiting existing data with the use of our theory and containing step-by-step description of programming operations which would simplify the life of possible exploiters.

Although main principles of this theory are elaborated, many important problems still remain to be solved. Let us outline some of them. As far as the Raman experiment
in  nanopowders is concerned, it would be very interesting to extend the theory to elongated particles and to the closely related problem of nanopowders with multi-parametric distribution functions. Furthermore, in present research we concentrated on intrinsic for nanoparticles mechanisms of disorder. Investigation of such extrinsic mechanisms of optical phonon line broadening as solvent or crystalline matrix impact as well as contact with other particles in a powder are of high interest. Next, extension of continuous EKFG treatment  onto anisotropic crystals requires more sophisticated tensor modification of the EKFG method itself. Moreover, the analysis of Raman scattering in polar crystals within the framework of the same theory as it is done for nonpolar ones would essentially extend the range of applicability of theory. Experimental search and proper theoretical description of surface (surface phonons) and surface-bulk mixed (i.e., breathing) modes in nanoparticles which should create their own Raman peaks is one more task for our method, the latter phenomena might be very helpful for testing the nanoparticle surface.

Considering possible extensions and generalizations of the theory of optical phonon line broadening in particles not necessarily appealing to Raman experiment we should mention the intriguing task to include into the theory anharmonicity-induced processes of inelastic phonon scattering by each other which would lead to temperature dependence of the phonon linewidth (by the way, seen also in Raman experiment \cite{chaigneau2012laser}). In the meantime, the lifetimes of optical phonons
(stemming from both inelastic and elastic processes) are a subject of extensive experimental investigation in other confined systems such as quantum dots and short nanotubes. Our treatment of inelastic phonon rates adjusted for Raman experiment in nanoparticles  could be applied for these systems without significant revision.

Addressing possible generalizations of the present theory for Raman processes due to different (non-phonon) excitations one should mention first the case of strongly disordered (probably, amorphous) particles where the notion of propagating phonon modes is meaningless, and the vibrational dynamics is due to ``fractons''~\cite{Nakayama1994}. We believe that the Raman experiment in amorphous particles is better described by these excitations. Al last, it would be very beneficial to develop  EKFG-like theory for  Raman (or Mandelstam-Brillouin) scattering due to magnons in magnetically ordered particles where Bloch equations should replace phonon equations of motion.


\acknowledgments

The authors are thankful to Igor Gornyi for valuable comments.
This work is supported by the Russian Science Foundation (Grant No. 19-72-00031).

\bibliography{raman}

\end{document}